\documentclass[conference]{IEEEtran}
\PassOptionsToPackage{hyphens,spaces,obeyspaces}{url}
\usepackage[dvipsnames,table,xcdraw]{xcolor}
\usepackage{amsmath}
\usepackage{xspace}
\usepackage{color}
\usepackage{listings}
\usepackage{amssymb,wasysym,pifont}
\usepackage{graphicx}
\usepackage{subcaption}
\usepackage{hyphenat}
\usepackage[frozencache,cachedir=.]{minted}
\usepackage{soul}
\usepackage{placeins}
\usepackage{pgf}
\usepackage{import}
\usepackage{tikz}
\usepackage{tabularx,booktabs,multicol,multirow,colortbl,threeparttable}
\usepackage[boxruled]{algorithm2e}
\usepackage[inline]{enumitem}
\usepackage{etoolbox}
\usepackage{adjustbox}
\usepackage{tcolorbox}
\usepackage[colorlinks=true]{hyperref}
\usepackage[nameinlink,capitalise]{cleveref}
\usepackage[numbers,sort]{natbib}
\usepackage[nobiblatex]{xurl}

\setminted[c]{breaklines,xleftmargin=2em,linenos,tabsize=4,obeytabs,frame=single,framesep=5pt}
\setminted[gas]{breaklines,xleftmargin=2em,linenos,tabsize=4,obeytabs,frame=single,framesep=5pt}

\newcommand{\parhead}[1]{\vspace{1pt plus 1pt minus 0.5pt}\par\noindent\textbf{#1}\hspace{.75em plus .5em minus .5em}}

\newcommand{\attackname}{CacheOut\xspace}

\newcommand{\yes}{{\ttfamily \color{ForestGreen}\checkmark}\xspace}%
\newcommand{\no}{{\ttfamily \color{BrickRed}\ding{55}}\xspace}%
\newcommand{\stripe}{\rowcolor{blue!5}}

\newcommand{\thread}{hyper-thread\xspace}
\newcommand{\threads}{{\thread}s\xspace}
\newcommand{\threading}{{\thread}ing\xspace}
\newcommand{\threaded}{{\thread}ed\xspace}

\newcommand{\cacheattack}[2]{{\tt\sc #1+\allowbreak #2}\xspace}
\newcommand{\flushreload}{\cacheattack{Flush}{Reload}}
\newcommand{\primeprobe}{\cacheattack{Prime}{Probe}}

\newcommand{\clflush}{{\ttfamily clflush}\xspace}

\newcommand{\xbegin}{{\ttfamily xbegin}\xspace}
\newcommand{\xend}{{\ttfamily xend}\xspace}

\newcommand{\verw}{{\ttfamily verw}\xspace}
\newcommand{\movq}{{\ttfamily movq}\xspace}
\newcommand{\eldu}{{\ttfamily eldu}\xspace}
\newcommand{\ewb}{{\ttfamily ewb}\xspace}
\newcommand{\cpuid}{{\ttfamily cpuid}\xspace}

\usepackage[symbol]{footmisc}

\newcommand{\ldflush}{{\ttfamily MSR\_IA32\_FLUSH\_CMD}\xspace}

\newcommand{\tsxctrl}{{\ttfamily MSR\_IA32\_TSX\_CTRL}\xspace}

\newcommand{\hrtick}{{\ttfamily hrtick}\xspace}
\newcommand{\yield}{{\ttfamily sched\_yield()}\xspace}

\crefname{figure}{Figure}{Figures}

\begin{document}
\date{}
\title{\attackname: Leaking Data on Intel CPUs \\ via Cache Evictions}

\include{utils}

\makeatletter
\newcommand{\linebreakand}{%
\end{@IEEEauthorhalign}
\hfill\mbox{}\par
\mbox{}\hfill\begin{@IEEEauthorhalign}
}
\makeatother

\author{ 
{\rm Stephan van Schaik* }\\
University of Michigan \\
\rm \url{stephvs@umich.edu}\\
\and
{\rm Marina Minkin}\\
University of Michigan\\
\rm \url{minkin@umich.edu}
\and
{\rm Andrew Kwong}\\
University of Michigan\\
\rm \url{ankwong@umich.edu}
\linebreakand
{\rm Daniel Genkin}\\
University of Michigan \\
\rm \url{genkin@umich.edu}
\and
{\rm Yuval Yarom}\\
University of Adelaide {\small and} Data61\\
\rm \url{yval@cs.adelaide.edu.au}
}

\maketitle

\begin{abstract}
Recent transient-execution attacks, such as RIDL, Fallout, and ZombieLoad, demonstrated that attackers can leak information while it transits through microarchitectural buffers.
Named Microarchitectural Data Sampling (MDS) by Intel, these attacks are likened to ``drinking from the firehose'', as the attacker has little control over what data is observed and from what origin.
Unable to prevent the buffers from leaking, Intel issued countermeasures via microcode updates that overwrite the buffers when the CPU changes security domains.

In this work we present \attackname, a new microarchitectural attack that is capable of bypassing Intel's buffer overwrite countermeasures.
We observe that as data is being evicted from the CPU's L1 cache, it is often transferred back to the leaky CPU buffers where it can be recovered by the attacker.
\attackname improves over previous MDS attacks by allowing the attacker to choose which data to leak from the CPU's L1 cache, as well as which part of a cache line to leak.
We demonstrate that \attackname can leak information across multiple security boundaries, including those between processes, virtual machines, user and kernel space, and from SGX enclaves.
\end{abstract}

\section{Introduction}
\label{sec:intro}
\footnotetext[1]{Work partially done while author was affiliated with Vrije Universiteit Amsterdam.}

In 2018 Spectre~\cite{Kocher2018spectre} and Meltdown~\cite{meltdown} left an everlasting impact on the design of modern processors.
Speculative and out-of-order execution, which were considered to be harmless and important CPU performance features, were discovered to have severe and dangerous security implications.
While the original Meltdown and Spectre  works focused on breaking kernel-from-user and process-from-process isolation, many follow-up works have demonstrated the dangers posed by uncontrolled speculation and out-of-order execution.
Indeed, these newly-discovered \emph{transient-execution attacks} have been used to violate numerous security domains, such as Intel's Secure Guard Extension (SGX)~\cite{foreshadow}, virtual machine boundaries~\cite{foreshadow-ng}, AES
hardware accelerators~\cite{stecklina2018lazyfp} and others~\cite{canella2019systematic,chen2019sgxpectre,horn2018spec4,kiriansky2018speculative,Kocher2018spectre, koruyeh2018spectre,maisuradze2018ret2spec,smotherspectre,islam2019spoiler,swapgs}.

More recently, the security community uncovered a deeper source of leakage: internal and mostly undocumented CPU buffers.
With the advent of Microarchitectural Data Sampling (MDS) attacks~\cite{ridl, fallout, zombieload}, it was discovered that the contents of these buffers can be leaked via assisting or faulting load instructions, bypassing the CPU's address and permission checks.
Using these techniques, an attacker can siphon off data as it appears in the buffer, bypassing all previous hardware and software countermeasures and again breaking nearly all hardware-backed security domains.

Responding to the threat of unconstrained data extraction, Intel deployed countermeasures for blocking data leakage from internal CPU buffers.
For older hardware, Intel augmented a legacy x86 instruction, \verw, to overwrite the contents of the leaking buffers.
This countermeasure was subsequently deployed by all major operating system vendors, performing buffer overwrite on every security domain change.
In parallel, Intel launched the new Whiskey Lake architecture, which is designed to be resistant to MDS attacks~\cite{ridl, fallout, zombieload}.

While the intuition behind the buffer overwrite countermeasure is that  an attacker cannot recover buffer information that is no longer present, previous works~\cite{zombieload,ridl} already report observing some residual leakage despite buffer overwriting.
Thus, in this paper we ask the following questions:

\smallskip
{\centering \emph{Are buffer overwrites sufficient to block MDS-type attacks? How can an adversary exploit the the buffers in Intel CPUs despite their content being properly overwritten?}
}
\smallskip

Moreover, for Whiskey Lake machines, we note that the nature of Intel's hardware countermeasures is not documented, requiring users to blindly trust Intel that MDS has been truly mitigated.
Thus, we ask the following secondary questions:

\smallskip
{\centering \emph{Are Whiskey Lake machines truly resistant to MDS attacks? How can an attacker leak data from these machines despite Intel's hardware countermeasures?}
}

\subsection{Our Contribution}

\noindent Unfortunately, we show that ad-hoc buffer overwrite countermeasures as well as Intel's hardware mitigations are both insufficient to completely mitigate MDS-type attacks.
More specifically, we present \attackname, a transient-execution attack that is capable of bypassing Intel's buffer overwriting countermeasures as well as leak data on MDS-resistant Whiskey Lake CPUs.
Moreover, unlike prior MDS works, \attackname allows the attacker to select which cache sets to read from the CPU's L1 Data cache, as opposed to being limited to data present in the 12 entries of the line fill buffers.
Next, because the L1 cache is often not flushed on security domain changes, \attackname is effective even in the case without \threading, where the victim always runs sequentially to the attacker.
Finally, we show that \attackname is applicable to nearly all hardware-backed security domains, including process and kernel isolation, virtual machine boundaries, and the confidentiality of SGX enclaves.

\parhead{A New Type of LFB Leakage.}
We begin by observing that Intel's MDS countermeasures (e.g., the \verw instruction) do not address the root cause of MDS.
That is, even after Intel's microcode updates, it is still possible to use faulting or assisting loads to leak information from internal CPU buffers.
Instead, Intel's \verw instruction overwrites all the stale information in these buffers, sanitizing their contents. We further note previous observations by ZombieLoad and RIDL~\cite{ridl,zombieload}, which report residual leakage from the Line Fill Buffers (LFBs) on MDS-vulnerable processors despite the \verw mitigation.

At a high level, the line fill buffers are intended to provide a non-blocking operation of the L1-D cache by handling data retrieval from lower levels of the memory architecture when a cache miss occurs~\cite{us5671444,us5680572}.
Despite their intended role of \textit{fetching} data into the L1-D, we empirically find that on Intel CPUs there exists an undocumented path where data \textit{evicted} from the L1-D cache occasionally ends up inside the LFB.

\parhead{\attackname Overview.}
Exploiting this path, in this paper we show a new technique where we first evict data from the L1-D, and subsequently use a faulting or assisting load to recover it from the LFB. This technique has two important security implications.
First, in contrast to prior MDS attacks which can only access information that transits through the CPU's internal buffers, \attackname can leak information present in the entire L1-D cache by simply evicting it.
Next, we demonstrate that this information path between L1-D evictions and the LFB has devastating consequences on countermeasures that are rely on flushing buffers on security domain changes.
In particular, using Intel's \verw instruction on does not protect against \attackname,  because the transfer of evicted data from the L1 cache to the LFB occurs well after the context switch and the completion of the associated \verw instruction.

\parhead{Attacking Whiskey Lake Processors.}
We note that the information path from the L1-D to the LFB exists on Intel's latest Whiskey Lake processors, which protect against MDS attacks via hardware countermeasures as opposed to the \verw instruction.
In addition to not leaking from internal CPU buffers, these machines also contain hardware mitigations against prior Meltdown and Foreshadow/L1TF attacks which leak information from the L1 cache.
Thus, to the best of our knowledge, \attackname is the first demonstration of a successful transient-execution attack on Whiskey Lake CPUs, which do not directly leak either from the LFB nor from the L1 cache.

\parhead{Leakage Amount.}
As noted above, the presence of data leakage despite using the \verw instruction has been previously observed by both the RIDL~\cite{ridl} and the ZombieLoad~\cite{zombieload} teams.
RIDL does not report any rates but only shows leakage via statistical significance.
ZombieLoad reports a troubling but insignificant amount of leakage, around 0.1\,B/s~\cite[Section~7]{zombieload}.
In this work we show that the leakage is significantly higher, peaking out at around 2.85\,KiB/s.

\parhead{Controlling What to Leak.}
Our technique of forcing L1 eviction also allows us to select the data to leak from the victim's address space.
Specifically, the attacker can force contention on a specific cache set, causing eviction of victim data from this cache set, and subsequently use the TAA attack~\cite{deep-dive:taa} to leak this data after it transits through the LFB.
To further control the location of the leaked data, we observe that the LFB seems to have a \emph{read offset} that controls the position within the buffer from which a load instruction reads.
We observe that some faulting or assisting loads can use stale offsets from \emph{subsequent} load instructions.
Combined with cache evictions, this allows us to control the 12 least significant bits of the address of the data we leak.

Finally, by repeating this technique across all 64 L1-D cache sets, \attackname is able to dump entire 4\,KiB pages from the victim's address space, recoving data as well as the positions of data pieces relative to each other.
This significantly improves over previous MDS attacks which can only recover information as it transits through the LFB without its corresponding location; TAA has the additional limitation of being able to read only the first 8 bytes of every cache line present in the LFB, leaving the other 56 bytes inaccessible.

\parhead{Attacking Loads.}
Cache eviction is useful for leaking data from cache lines \emph{modified} by the victim. This is because the victim's write marks the corresponding cache line as dirty, forcing the CPU to move the data out of the cache and to the LFB. It does not, however, allow us to leak data that is only \emph{read} by the victim, since this data is not written back to memory and does not occupy a buffer when evicted from the L1 cache. We overcome this by evicting the victim's data from the L1 before the victim has a chance to read it. This induces an L1 cache miss, which is served via the LFB. Finally, we use an attacker process running on the same physical core as the victim to recover the data from the LFB.

\parhead{Attacking Process Isolation.}
We show that \attackname has severe implications for OS-enforced process security boundaries, as it allows unprivileged users to read information belonging to other victim processes, thereby breaching their confidentiality.
We demonstrate this risk by implementing attacks on private data across processes in different security domains.
Targeting OpenSSL's AES operations, we successfully recover secret keys and plaintext data in both the scenarios with and without \threading.
We also developed attacks for both recovering OpenSSL RSA private keys and stealing the secret weights from a FANN artificial neural network.

\parhead{Attacking the Linux Kernel.}
Beyond proof-of-concept exploits, we also demonstrate highly practical attacks against the Linux kernel, all mounted from unprivileged user processes.
By taking advantage of \attackname's cache line selection capabilities, we are able to completely derandomize Kernel Address Space Layout Randomization (KASLR) in under a second.
Furthermore, we demonstrate extraction of stack canaries from the kernel.
To the best of our knowledge, this is the first demonstration of acquiring this information via a MDS-type transient-execution attack.

\parhead{Attacking Intel's Secure Guard Extensions (SGX).}
We demonstrate that \attackname can dump the contents of SGX enclaves. We show the recovery of an image from SGX enclaves as well as of EPID keys in debug mode.
These attacks are performed on a fully updated Whiskey Lake CPU, which is resistant to previous MDS attacks, including Fallout~\cite{fallout}, ZombieLoad~\cite{zombieload} and RIDL~\cite{ridl} and TSX Asynchronous Abort (TAA)~\cite{deep-dive:taa}.
In particular, this implies that Intel's current hardware mitigations for SGX are insufficient, allowing an attacker to breach the confidentiality of SGX enclaves.

Moreover, \attackname can dump the contents of an enclave without requiring it to perform any operation or even execute at all.
Instead, we directly dump the memory content of the victim enclave while the enclave is idle.
Thus, our attack bypasses all software-based SGX side-channel defenses such as constant-time coding and others~\cite{shih2017t, chen2018racing, oleksenko2018varys,fu2017s,sasy2017zerotrace} which rely on the enclave executing code for its protection.

\parhead{Attacking Virtual Machines.}
Another security domain we explore is the isolation of different virtual machines running on the same physical core.
We show that \attackname is effective at leaking data from both virtual machines and from hypervisors.
Experimentally evaluating this, we completely derandomize the Address Space Layout Randomization (ASLR) used by the hypervisor and recover AES keys from another VM.

\parhead{Avoiding Hyper-Threading.}
While \attackname is most effective across \threads, we can nonetheless use it to recover information in a time-shared environment, with \threading being disabled, even in the presence of the \verw countermeasure.
The core failure is that the \verw instruction only flushes the internal CPU buffers, and not the L1 cache.
Thus, an attacker can evict cached data left by the victim and subsequently recover it from the leaky line fill buffer.
Finally, \attackname is able to defeat the hardware countermeasures on Whiskey Lake CPUs, both with and without \threading.

\parhead{Summary of Contributions.}
In this paper we make the following contributions: 

\begin{itemize}[leftmargin=*, nolistsep]
\item We present \attackname, the first transient-execution attack that can leak across arbitrary address spaces while still retaining fine grained control over what data to leak.
Moreover, unlike other MDS-type attacks, \attackname cannot be mitigated by simply overwriting the contents of internal CPU buffers between context switches, even when \threading is disabled.
\item We demonstrate the effectiveness of \attackname in violating process isolation by recovering AES and RSA keys as well as plaintexts from an OpenSSL-based victim.
\item We demonstrate practical exploits for completely derandomizing Linux's kernel ASLR, and for recovering secret stack canaries from the Linux kernel.
\item We demonstrate how \attackname violates isolation between two virtual machines running on the same physical core.
\item We breach SGX's confidentiality guarantees by reading out the contents of an SGX enclave and recovering the machine's attestation keys from a fully updated system.
\item We demonstrate that some of the latest Intel CPUs are still vulnerable, despite all of the most recent patches and mitigations. In particular, to the best of our knowledge, \attackname is the first transient-execution attack to break Intel's MDS-resistant Whiskey Lake architecture.
\item We discuss why current transient-execution attack mitigations are insufficient, and offer suggestions on what countermeasures would effectively mitigate \attackname.
\end{itemize}

\subsection{Current Status and Disclosure}

Van~Schaik et al.~\cite{ridl} note the relationship between cache evictions and MDS attacks.
The first author and researchers from VU Amsterdam notified Intel about the findings contained in this paper during October 2019 Intel acknowledged the issue and assigned \texttt{CVE-2020-0549}, referring to the issue as L1 Data Eviction Sampling (L1DES) with a CVSS score of 6.5 (medium).
Intel has also informed that L1DES has been independently reported by researchers from TU Graz and KU Leuven.

\parhead{Current Status.}
In November 2019, after our initial disclosure of \attackname, Intel attempted to mitigate TSX Asynchronous Abort (TAA)~\cite{deep-dive:taa}, a variant of MDS which allows an attacker to leak information from internal CPU buffers.
Consequently, in November 2019 Intel published microcode updates that enable turning off Transactional Memory Extension (TSX) on CPUs made after Q4 2018.
These have been deployed by OS vendors, preventing some variants of \attackname on these machines.
However, for SGX, a malicious OS can always re-enable TSX.
As we show in this paper, this results in a loss of confidentiality due to our breach of Intel's TAA countermeasures for protecting SGX.

Next, we note that the majority of deployed Intel hardware is older, and was released prior to Q4 2018.
For these systems, TSX is enabled by default at the time of writing, leaving them vulnerable to all variants of \attackname.
Finally, Intel had indicated that microcode updates mitigating the root cause behind \attackname will be published 
on June 9th, 2020.
We recommend these be installed on all affected Intel platforms to properly mitigate \attackname.

\section{Background}
\label{sec:background}

\subsection{Caches}
\label{sec:caches}

To bridge the performance gap between the CPU and main memory, processors contain small buffers called \emph{caches}.
These exploit locality by storing frequently and recently used data to hide the access latency of main memory.
Modern processors typically include multiple caches.
In this work we are mainly interested in the L1-D cache, which is a small cache that stores data the program uses.
A multi-core processor typically has one L1-D cache in each processor core.

\parhead{Cache Organization.}
Caches generally consist of multiple cache sets that can host up to a certain number of cache lines or \emph{ways}.
Part of the virtual or physical address of a cache line maps that cache line to its respective cache set, where \emph{congruent} addresses are those that map to the same cache set.

\parhead{Cache Attacks.}
An attacker can infer secret information from a victim in a shared physical system such as a virtualized environment by monitoring the victim's cache accesses.
Previous work proposed many different techniques to perform cache attacks, the most notable among them being \flushreload and \primeprobe.

\flushreload attacks~\cite{gullasch2011cache,flush+reload} work with shared memory at the granularity of a cache line.
The attacker repeatedly flushes a cache line using a dedicated instruction, such as \clflush, and then measures how long it takes to reload the cache line.
A fast reload time indicates that another process brought the cache line back into the cache.

\primeprobe~\cite{osvik,percival,shared-cache-attack,llc-practical,hires-llc-attack} attacks, on the other hand, work without shared memory, but only at the granularity of a cache set.
The attacker repeatedly accesses an \emph{eviction set}---a set of congruent memory addresses that fills up an entire cache set---while  measuring how long that takes.
As the attacker repeatedly fills up the entire cache set with their own cache lines, the access time is generally
low.
However, when another process accesses a memory location in the same cache set, the access time becomes higher because the victim's cache line replaces one of the lines in the eviction set.

\subsection{Microarchitectural Buffers}
\label{sec:uarch-buffers}

\begin{figure}[t]
\centering
\includegraphics[width=\linewidth]{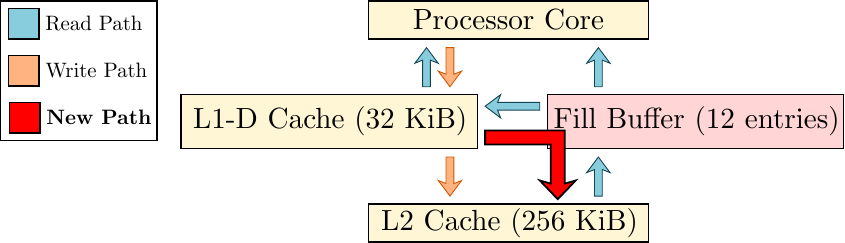}
\caption{The data paths within the CPU core, with the paths for loads marked in blue, the path for stores in orange, and the new undocumented path that we uncovered marked in red. }
\label{fig:datapath}
\end{figure}

In addition to caches, modern processors contain multiple microarchitectural buffers that are used for storing data in-transit.
In this work we are mainly interested in the \emph{Line Fill Buffers}, depicted in \cref{fig:datapath}, which handle data transfer between the L1-D cache, the L2 cache, and the core.

\parhead{Non-Blocking L1-D Cache Misses.}
One purpose of the line fill buffers is to enable non-blocking operation mode for the L1-D cache~\cite{us5671444,us5680572} by handling the retrieval of data from lower levels of the memory architecture when a cache miss occurs.
Specifically, when the processor services a load instruction, it consults both the LFBs and the L1-D cache in parallel.
If the data is available in either component, the processor forwards the data to the load instruction.
Otherwise, the processor allocates an entry in the LFB to keep track of the address, and issues a request for the data from the L2 cache.
When the data arrives, the processor forwards it to all pending loads.
The processor may also allocate an entry for the data in the L1-D cache, where it is stored for future use.

\parhead{A New Data Path.}
As mentioned above, while the LFB is responsible for handling data coming into the L1-D cache, we empirically demonstrate the existence of an undocumented data path between L1-D evictions and the LFB (marked in red in~\cref{fig:datapath}).
We then exploit this path by causing L1-D evictions and subsequently leak the evicted data from the LFB.
In addition to bypassing the \verw instruction by moving data into the LFB after the \verw-induced buffer overwrite, we also show that this path exists on MDS-resistant Whiskey Lake machines, making these vulnerable to \attackname.

\subsection{Speculative and Out-of-Order Execution}

Modern processors try to predict future instructions and execute instructions as soon as the required data is available, rather than following the strict order stipulated by the program.
Because the exact sequence of future instructions is not always known in advance, the processor may sometimes execute \emph{transient} instructions that are not part of the nominal program execution.
This can occur, for example, when the processor mispredicts the outcome of a branch instruction and executes instructions following the wrong branch.
When the processor determines that an instruction is transient, it drops all of the results of the instruction instead of committing them to the architectural state.
Consequently, transient instructions do not affect the architectural state of the processor.

\subsection{Transient-Execution Attacks}

Because transient instructions are not part of the nominal program order, they may sometimes process data that is not accessible in nominal program order.
In recent years, multiple \emph{transient-execution attacks} have demonstrated the possibility of leaking such data~\cite{canella2019systematic, chen2019sgxpectre,horn2018spec4, kiriansky2018speculative,Kocher2018spectre, koruyeh2018spectre, maisuradze2018ret2spec}.
In a typical attack, the attacker induces speculative execution of transient instructions that access secret data and leak it back to the attacker.
Because the instructions are transient, they cannot transmit the secret data via the architectural state of the processor.
However, execution of transient instructions can modulate the state of microarchitectural components based on the secret data.
The attacker then probes the state of the microarchitectural component to determine the secret data.

Most published transient-execution attacks use a \flushreload-based covert channel for sending the data.
In a typical attack, the attacker maintains a \emph{probing} array consisting of 256 distinct cache lines.
The attacker flushes all of these cache lines from the cache before causing speculative execution of the attack \emph{gadget}.
Transient instructions in the attack gadget access a secret data byte, and use it to index a specific cache line in the probing array, bringing the line into the cache.
The attacker then performs the reload step of the \flushreload attack to identify which of the probing array's cache lines is in the cache, revealing the secret byte.

\subsection{RIDL, ZombieLoad, Medusa vs. CacheOut}

Several prior works explore leakage from internal CPU buffers.
These include RIDL~\cite{ridl}, ZombieLoad~\cite{zombieload}, Fallout~\cite{fallout}, and Medusa~\cite{medusa}, collectively known as \emph{MDS attacks}.

\parhead{RIDL.}
RIDL~\cite{ridl} analyzes the Line Fill Buffers (LFBs) and the Load Ports.
Focusing mainly on the case with \threading, the work shows that faulting loads can be served from these buffers, bypassing any address and permission checks.
This allows an attacker to use the sibling core to siphon off data as it appears in the buffer, compromising the confidentiality of nearly all hardware-backed security domains.

However, while RIDL conjectures that data evicted from the L1 cache is moved into the leaky LFB, and even shows some statistical evidence of such leakage, it does not study the security implications of this issue nor of Intel's MDS buffer flush countermeasures.
Moreover, RIDL also lacks control over what data the attacker is leaking, and instead relies on averaging techniques to filter the data from the acquired noise.
Finally, the lack of control over what can be leaked from the L1 cache implies that RIDL only demonstrates attacks in the \threaded case, where the attacker siphons off data from the LFB as the victim accesses it.

\parhead{ZombieLoad.}
ZombieLoad~\cite{zombieload} also analyzes leakage from the LFBs.
Extending RIDL's findings to loads that require microcode assists, ZombieLoad shows that leakage exists even without
using faulting loads.
ZombieLoad also demonstrates LFB leakage from the Cascade Lake architecture, that Intel claims to be the first MDS-resistant architecture.

Similarly to RIDL, ZombieLoad mentions the possibility of leakage via L1 evictions to the LFB.
However, ZombieLoad proceeds to argue that the leakage is negligible, limited to 0.1 bytes per second.
ZombieLoad also suffers from limitations similar to RIDL with regards to the attacker's ability to control the leakage, resorting to Domino-bytes averaging techniques for data processing with the attacker and victim running on different threads of the same physical core.
While the ZombieLoad paper mentions kernel attacks in the case without \threading, Section 6.5 in~\cite{zombieload} only demonstrates attacks using artificially inserted kernel gadgets, at a rate of 10 seconds per byte.
Finally, while ZombieLoad does mention the possibility of hypervisor and cross-VM leakage, \citet{zombieload} only demonstrate a cross-VM covert channel.

\parhead{Medusa.}
In concurrent independent work, \citet{medusa} presented Medusa, a variant of ZombieLoad that recovers information from write-combining (WC) operations~\cite{intelCopying}, for which the LFB is responsible on Intel CPUs~\cite[vol. 3 pg. 6-38]{intel-sdm}.
By focusing on leakage from write combining done in the LFB during \texttt{rep mov} and \texttt{rep stos} operations, Medusa is able to obtain a cleaner LFB leakage signal, as it avoids recovering values from other memory operations.
Finally, as OpenSSL uses fast memory copying to copy RSA keys, and the kernel to transfer data, Medusa demonstrated the recovery of such data across \threads.

However, Medusa is (intentionally) limited to only recovering values during write combining, and is unable to recover leakage from other memory operations.
Being a variant of ZombieLoad, the attacker has no knowledge or control over the exact offsets, and can only partially sample the leaked data.
This results in slow leakage rates of 12 B/s for kernel data, and the need for Domino-bytes signal averaging.
For unstructured data (e.g., RSA keys), a 400 CPU hour lattice attack~\cite{coppersmith1997small} is needed to recover the 1024-bit RSA key from the raw leakage, which is obtained during a 7 minutes measurement phase.

\parhead{CacheOut.}
In this work, we also focus on leakage from Intel's LFB.
However, unlike RIDL, ZombieLoad, and Medusa we do not wait for the information to become available in the LFB, and instead use cache evictions to actively move it to the leaky LFB.
We show that the leakage is far greater than the 0.1B/s conjectured by ZombieLoad, peaking out at 2.85KiB/s.
Next, we show that by using cache evictions the attacker can choose what information he is interested in leaking, thus avoiding the need to use noise-averaging techniques. Ironically, in addition to bypassing Intel's \verw countermeasure that overwrites the contents of the leaky buffers, we go a step further and show how \verw can actually be used to improve our attack's leakage rate.
Furthermore, we show the effectiveness of \attackname in breaking the isolation between processes, VMs, hypervisors, and SGX enclaves.
Finally, we show that MDS attacks are still effective on Intel's latest MDS-resistant Whiskey Lake CPUs.

\subsection{TSX Asynchronous Abort}

Some contemporary Intel processors implement memory transactions through the \emph{Transactional Synchronization Extensions} (TSX).
As part of the extension, TSX offers the \xbegin and \xend instructions to mark the start and end of a transaction, respectively.
These instructions form a transaction where either all of them execute to completion or none of them at all.
All of the transaction's instructions execute speculatively, but are only committed if execution reaches the \xend instruction.
If during the execution of a transaction any instruction in the transaction faults, the transaction is aborted and all of the instructions in the transaction are dropped.

The Intel manual states that there are CPU implementations where the \clflush instruction may always cause a transactional abort with TSX \cite[vol. 2A, pg. 3-139--3-142]{intel-sdm}.
TSX Asynchronous Abort (TAA) \cite{deep-dive:taa,ridl,zombieload} exploits this behavior in a transient-execution attack by flushing cache lines before running a transaction that attempts to load data from the flushed cache line.
Reading from the flushed line aborts the transaction.
However, before the transaction aborts, the processor allocates an LFB entry for the load instruction.
When the transaction aborts, the load instruction is allowed to proceed speculatively with data from the LFB.
Since the load does not complete successfully, the load proceeds with remnants from a previous memory access, allowing the attacker to sample LFB data~\cite{deep-dive:mds,deep-dive:taa}.
We refer the reader to Appendix~\ref{app:taa} for a TAA code example.

\section{CPU Mitigations and Threat Model}
\label{sec:threat}

\newcolumntype{L}{>{$}l<{$}}
\newcolumntype{C}{>{$}c<{$}}
\newcolumntype{R}{>{$}r<{$}}
\begin{table*}[t]
\centering
\small
\begin{tabular}{ lcc c ccccc }
  \toprule
  CPU & Year & CPUID && Meltdown & Foreshadow & MDS & TAA & \attackname \\
\midrule
  Intel Xeon Silver 4214 (Cascade Lake SP) & Q2 '19 & 50657 && \yes & \yes & \yes & \no & \no \\
\stripe
  Intel Core i7-8665U (Whiskey Lake) & Q2 '19 & 806EC && \yes & \yes & \yes & \no & \no \\
  Intel Core i9-9900K (Coffee Lake Refresh - Stepping 13) & Q4 '18 & 906ED && \yes & \yes & \yes & \no & \no \\
\stripe
  Intel Core i9-9900K (Coffee Lake Refresh - Stepping 12) & Q4 '18 & 906EC && \yes & \yes & \no & \no & \no \\
  Intel Core i7-8700K (Coffee Lake) & Q4 '17 & 906EA && \no & \no & \no & \no & \no \\
\stripe
  Intel Core i7-7700K (Kaby Lake) & Q1 '17 & 906E9 && \no & \no & \no & \no & \no \\
  Intel Core i7-7800X (Skylake X) & Q2 '17 & 50654 && \no & \no & \no & \no & \no \\
\stripe
  Intel Core i7-6700K (Skylake) & Q3 '15 & 506E3 && \no & \no & \no & \no & \no \\
  Intel Core i7-6820HQ (Skylake) & Q3 '15 & 506E3 && \no & \no & \no & \no & \no \\
\bottomrule
\end{tabular}
\vspace{-0.5em}
\caption{%
Countermeasures for transient execution attacks in Intel processors.
\yes and \no indicate the existence or absence of in-silicon countermeasure for the attack.}
\label{tab:eval}
\end{table*}

Since the discovery of Spectre~\cite{Kocher2018spectre} and Meltdown~\cite{meltdown}, there have been numerous works that exploit speculative and out-of-order execution to violate hardware-backed security domains~\cite{canella2019systematic,chen2019sgxpectre,horn2018spec4,kiriansky2018speculative,Kocher2018spectre,koruyeh2018spectre,maisuradze2018ret2spec}.
In response, recent Intel processor contain hardware-based countermeasures aimed at addressing these attacks.
\cref{tab:eval} summarizes these countermeasures in some recent Intel processors.
For processors that are not protected, Intel enabled some features that can be used to provide software-based protection.
We now describe these software-based countermeasures.

\parhead{Kernel Page Table Isolation (KPTI).}
Meltdown~\cite{meltdown,deep-dive:meltdown} shows that an attacker can bypass the protection of kernel memory.
The attack requires that the virtual address is present in the address space and that the data it refers to is present in the L1-D cache.
Thus, to mitigate Meltdown, operating systems deploy KPTI~\cite{gruss2017kaslr,Corbet17} or similar defenses that separate the kernel address space from the user address space, thereby rendering kernel addresses inaccessible to attackers.

\parhead{Flushing the L1-D Cache.}
KPTI alone soon turned out to be ineffective, as Foreshadow/L1TF~\cite{foreshadow,foreshadow-ng,deep-dive:l1tf} demonstrates that any data can be leaked from the L1-D cache by speculatively reading from the physical address corresponding with the data in L1-D cache.
Since the disclosure of Foreshadow, Intel CPUs introduced \ldflush to flush the L1-D cache upon a VM context switch.
When the MSR is unavailable, the Linux KVM resorts to writing 64\,KiB of data to 16 pages.
\footnote{\url{https://github.com/torvalds/linux/blob/aedc0650f9135f3b92b39cbed1a8fe98d8088825/arch/x86/kvm/vmx/vmx.c\#L5936}}

\parhead{Flushing MDS Buffers.}
Fallout~\cite{fallout}, RIDL~\cite{ridl}, ZombieLoad~\cite{zombieload}, and Medusa~\cite{medusa} show that attackers can leak data transiting through various internal microarchitectural buffers, such as the LFBs discussed in \cref{sec:background}.
To address these issues for older hardware, Intel provided microcode updates~\cite{mcu-guidance:mds} that repurpose the \verw instruction to flush these microarchitectural buffers by overwriting them.
The operating system has to issue the \verw instruction upon every context switch to effectively flush these microarchitectural buffers.

\parhead{The Whiskey Lake Architecture.}
In an attempt to mitigate MDS attacks in hardware, Intel also released the Whiskey Lake architecture, which contains hardware mitigations to MDS attacks (i.e., RIDL, Fallout, and ZombieLoad) as well as to Meltdown and Foreshadow/L1TF.
In particular, Whiskey Lake machines are not vulnerable to previous MDS techniques that leak from internal buffers or to older generation Meltdown/Foreshadow attacks which leak the contents of the L1-D cache.
As we show however, these machines are vulnerable to \attackname, making our attack the only attack currently capable of leaking the contents of L1-D on these machines.

\parhead{Threat Model.}
We assume that the attacker is an unprivileged user, such as a VM, or an unprivileged user process on the victim's system. For the victim, we assume an Intel-based system that has been fully patched against Meltdown, Foreshadow, and MDS either in hardware or software.
We further assume that there are no software bugs or vulnerabilities in the victim software, or in any support software running on the victim machine.
We also assume that TSX RTM is present and enabled.
Finally, we assume that the attacker can run on the same processor core as the victim.

\section{\attackname: Exploiting Cache Evictions}
\label{sec:overview}

We now start our exposure of \attackname and show that it can bypass Intel's buffer overwriting
countermeasures.
At a high level, \attackname forces contention on the L1-D cache to evict the data it targets from the cache.
We describe two variants.
First, in the case that the cache contains data modified by the victim, the contents of the cache line transits through the LFBs while being written to memory.
Second, when the attacker wishes to leak data that the victim does not modify, the attacker first evicts the data from the cache, and then obtains it when it transits through the line fill buffers to satisfy a concurrent victim read.
\cref{fig:overview} shows a schematic overview of these attacks, which we now describe.

\begin{figure}[t]
\centering
\includegraphics[width=\linewidth]{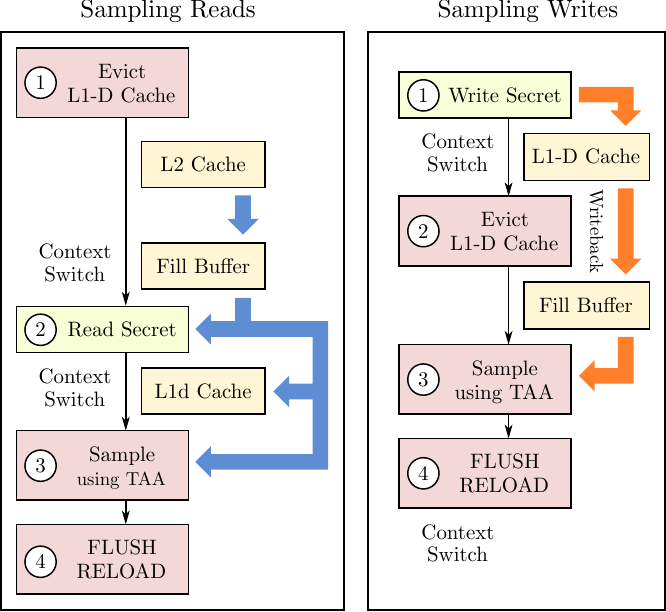}
\vspace{-1.2em}
\caption{%
Overview of how we use TAA to leak from loads and stores through {Fill Buffers}.
Victim activity, attacker activity, and microarchitectural effects are shown in green, red, and yellow respectively.
The context switches both illustrate the OS flushing the MDS buffers before switching to the other process as well as switching between actual \threads.}
\label{fig:overview}
\end{figure}

\parhead{Attacking Reads.}
The left part of \cref{fig:overview} shows our attacks on victim read operations.
We assume that the attacker has already constructed an eviction set for the cache set that contains the victim's data.
We further assume that the attacker and the victim run on two \threads of the same physical core.
For the attack, the attacker reads from all of the addresses in the eviction set (Step~1).
This loads the eviction set into the L1-D, evicting the victim's data.
Next, the attacker waits for the victim to access his data (Step~2).
This victim access brings the victim's data from the L2 cache into the line fill buffers, and subsequently to the L1 cache.
Finally, the attacker uses TAA (Step~3) to sample values from the line fill buffer and transmit them via a \flushreload channel (Step~4).

\parhead{Attacking Writes.}
\cref{fig:overview} (right) depicts our attack on victim write operations.
The attacker first waits until after the victim writes to a cache line (Step~1).
The attacker then accesses the corresponding eviction set, forcing the newly written data out of the L1-D cache and down the memory hierarchy (Step~2).
On its way to memory, the victim's data passes through the LFBs.
Thus, in Step~3, the attacker uses TAA to sample the buffer and subsequently uses \flushreload to recover the value (Step~4).
Finally, unlike the case of reads, this attack can be performed both with and without \threading.

\subsection{Exploiting L1-D Eviction}
\label{sec:l1d-eviction}

\parhead{Eviction Set Construction.} 
A precondition for \attackname is that the attacker is able to construct an eviction set for L1-D cache sets.
Recall that an eviction set is a collection of congruent addresses that all map to the same cache set.
On contemporary Intel processors, virtual addresses are used for addressing the L1-D cache.
Specifically, bits 6--11 of the virtual address are used to identify the cache set.
Consequently, the attacker can allocate eight 4\,KiB memory pages to cover the whole cache.
The attacker then constructs eviction sets from memory addresses with the same page offset.

\parhead{Measuring L1-D Eviction.}
To measure the number of accesses needed to evict the victim's line from the cache, we use a synthetic victim that repeatedly accesses the same cache set.
We test \attackname with varying eviction set sizes, and under three different attack scenarios.
\cref{fig:eviction} contains a summary of our results.
In the first scenario (left) the victim and the attacker time-share the same \thread.
As expected, when the eviction set contains eight addresses we get the best results, recovering the contents of the victim's cache line in 4.8\% of the cases.
Next, we note the decreased performance of the attack with other eviction set sizes (both smaller and larger sets).
We conjecture that small sets cannot evict the victim's element due to the cache's LRU replacement policy while larger eviction sets increase noise due to cache pressure. 

\begin{figure}[t]
\centering
\includegraphics[width=0.99\linewidth]{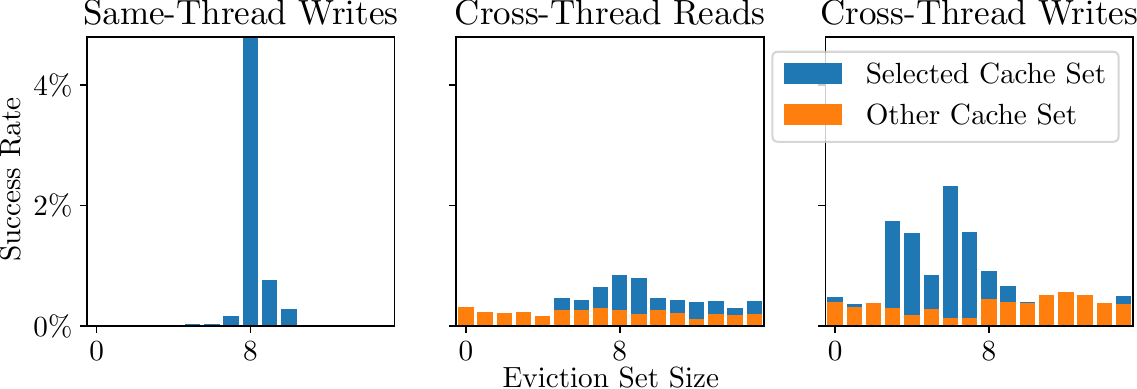}
\vspace{-1.5em}
\caption{%
Number of loads/stores required to evict the victim's cache line containing a fixed value.
The blue bars indicate how often we observe the correct value from the selected cache-line, while the orange bars indicate how often we observe data present in a different cache (due to noise).
Finally, we ran 10,000 iterations per tested set.}
\label{fig:eviction}
\end{figure}

We also test the cases across \threads, targeting the victim's memory reads (middle) and then victim writes (right).
While the results here are not as strong, the likelihood of getting the victim's data from the correct cache line is still higher than getting data from other cache lines.
For victim reads, we still get the best results with an eviction set of size eight while an eviction set of size six works best for writes.
We suspect that the cause is the increased L1-D contention due to having two active \threads.

\parhead{Measuring Data Selection.}
Demonstrating our ability to select which cache set to leak, we repeat the experiments of \cref{fig:eviction}, this time varying the cache set the victim uses and the cache set the attacker evicts.
As can be seen in \cref{fig:addrsel}, in all scenarios the attacker can target a specific cache set, correctly leaking its values, albeit with some noise for the case of cross-thread victim writes.
Finally, we note that this is a qualitative improvement over prior works such as RIDL~\cite{ridl}, ZombieLoad~\cite{zombieload} and Medusa~\cite{medusa}, as these are limited to leaking data already present in the 12 entries of the LFB, as opposed to leaking data from the entire L1-D cache.

\begin{figure}[t]
\centering
\includegraphics[width=0.99\linewidth]{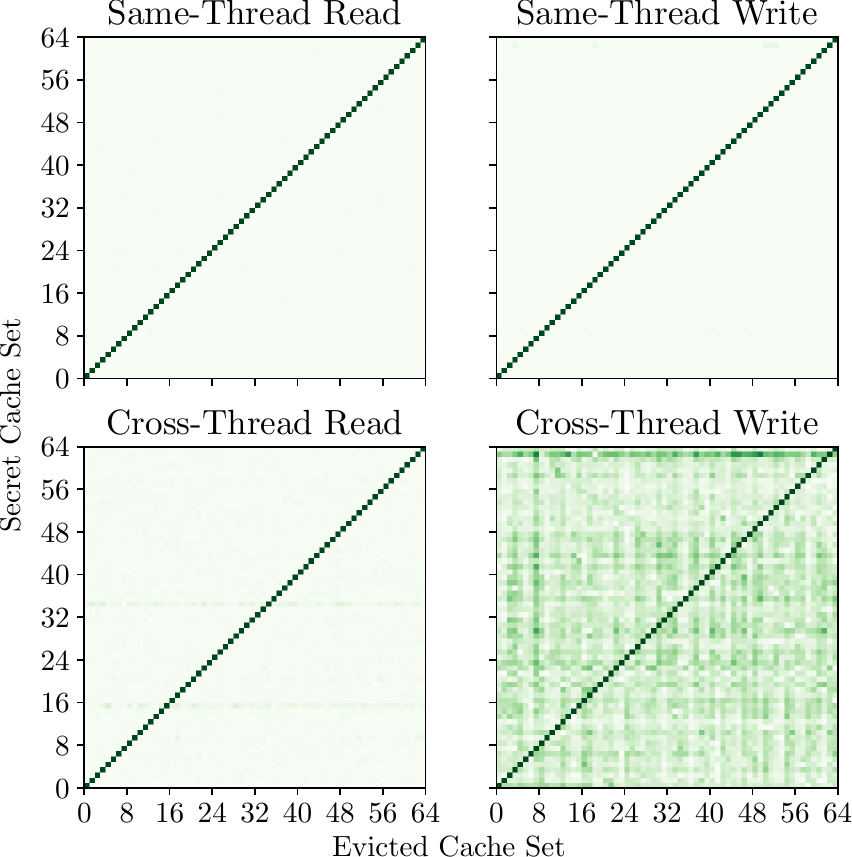}
\vspace{-1.2em}
\caption{%
The victim loads/stores a secret to every possible cache line (y-axis), while the attacker evicts every possible cache line (x-axis) to leak it.
We ran 10K iterations per test.}
\label{fig:addrsel}
\end{figure}

\subsection{Selecting Cache Line Offsets}
\label{sec:leaking-cache-lines}

\noindent 
So far we have shown how to control the cache set from which \attackname leaks data.
However, we note that like previous TAA attacks~\cite{ridl}, we still do not have control over the offset within the 64-bytes cache line from which we read.
In particular, TAA~\cite{ridl} is only able to leak the first 8 bytes out of every 64-byte cache line, leaving the other 56 bytes unreachable.

Tackling this limitation, we discovered that that the offset of load instructions that \emph{follow} the TAA attack also controls the cache line offset from which the TAA attack reads.
More specifically, \cref{lst:cacheout} shows our leakage primitive that allows us to control the offset from which we read data.
As we can see, the code is basically the same as the TAA leak primitive, but we added two \movq instructions in Lines 16--17.

\begin{listing}
\small
\begin{minted}{gas}
; %rdi = leak source
; %rsi = FLUSH + RELOAD channel
; %rcx = offset-control address
taa_sample:
	; Cause TSX to abort asynchronously.
	clflush (%rdi)
	clflush (%rsi)

	; Leak a single byte.
	xbegin abort
	movq (%rdi), %rax
	shl $12, %rax
	andq $0xff000, %rax
	movq (%rax, %rsi), %rax

	movq (%rcx), %rax
	movq (%rcx), %rax
	xend
abort:
	retq

\end{minted}
\vspace{-0.5em}
\caption{%
\attackname leak primitive.
\vspace{-1em}}
\label{lst:cacheout}
\end{listing}

\parhead{Analysing the \attackname Primitive.}
We note that the leakage in the \attackname primitive occurs in Line~11.
At a first glance it seems odd that later \movq instructions can affect the outcome of this instruction.
However, we note that  the \movq instructions we add do not depend on the outcome of the leaking \movq at Line~11.
Thus, due to out-of-order execution they can execute \emph{before} the instructions that precede them in program order.
We hypothesize that the line fill buffer has a \emph{read offset}, some internal state that determines the offset within buffer entries from which to read data.
This read offset gets reused by the leaking \movq when the transaction aborts, thereby allowing us to select the desired cache line offset.

\parhead{Reducing Noise and Data Stitching.}
Modern Intel CPUs typically employ two load ports, which allows them to execute two load instructions in parallel.
Exploiting this, in our attack we duplicated and interleaved the instructions from lines~11--14 in \cref{lst:cacheout}, such that they execute two load instructions in parallel on both load ports.
Compared to executing a single load instruction, we found that the strength of our signal doubles when both load instructions refer to the same offset.

Next, this technique also allows us to avoid using the Domino bytes method used in prior MDS works~\cite{ridl, zombieload, medusa}.
Instead, in our attack, we leak two consecutive data bytes at a time, where two consecutive attack iterations share a common data byte (at offset 2 in the first iteration and offset 1 in the second iteration).
Observing the leakage via the cache channel, we stitch together data that matches on the overlapping data byte as shown in \cref{fig:stitching}.

\begin{figure}[htb]
\centering
\includegraphics[width=\linewidth]{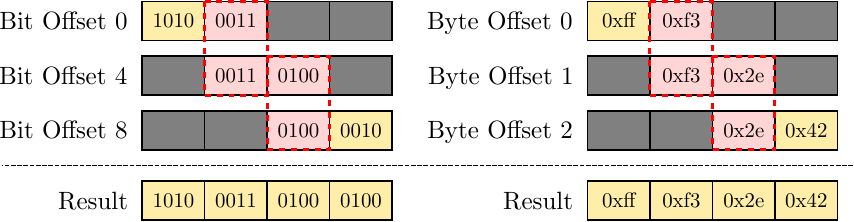}
\caption{%
On the left we show how the Domino attack samples a \emph{byte} at a time and uses four bits of every byte to stitch data together.
On the right we show our technique for CacheOut where the attacker samples \emph{two bytes} at a time and uses the leading and trailing byte to stitch data together, effectively doubling the attack's speed.}
\label{fig:stitching}
\end{figure}

\parhead{Evaluating Offset Selection.}
To evaluate our offset selection method, we use a victim process that chooses a byte offset and writes a secret value to this byte, setting the rest of the bytes in the same cache line to zero.
The attacker then tries to leak the secret from every possible byte offset from the victim's cache line.
As we can see in \cref{fig:byteleak}, we can successfully select the offset in the cache line from which we leak.
Next, combining this behavior with the L1-D cache set eviction method described in \cref{sec:l1d-eviction}, \attackname is equally effective against all addresses, and improves on prior MDS attacks by allowing the attacker to access any data located in the L1-D cache while being able to select the precise byte he is interested in leaking.

\parhead{Evaluating Leakage Amount.}
Finally, we evaluate the rate of information leakage resulting from exploiting targeted L1-D evictions into the leaky LFB.
Our victim writes some byte value to a known cache location, while the attacker running on the same physical core uses our address selection techniques in order to recover the victim's writes 10K times.
We distinguish between not leaking anything, leaking the correct value and leaking the incorrect value.
We find the leakage rate, i.e., how often we leak the correct value over a certain period of time, to be much larger than ZombieLoad's 0.1B/s, peaking out at 2.85KiB/s for reads and 2.38KiB/s for writes.

\begin{figure}[htb]
\centering
\includegraphics[width=.5\linewidth]{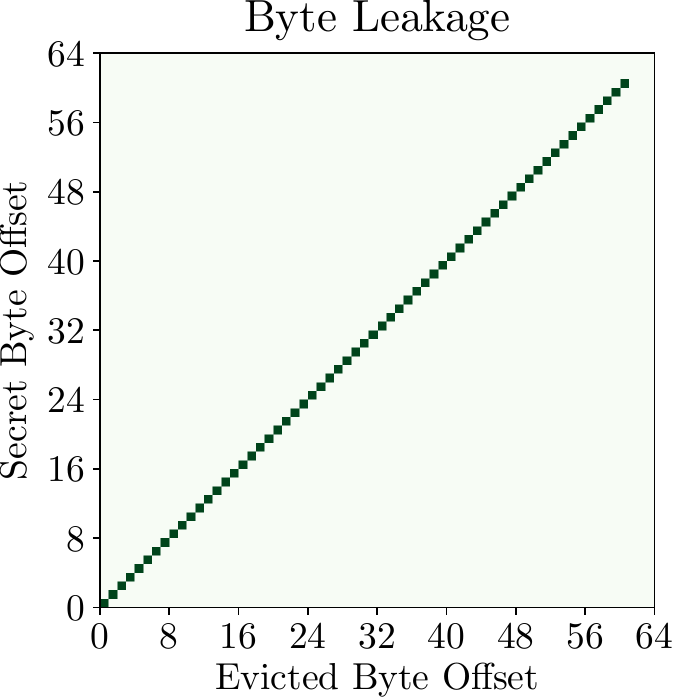}
\vspace{-0.3em}
\caption{%
The victim loads/stores a secret byte to every possible offset within a fixed cache line (y-axis), while the attacker tries to leak from every possible byte offset (x-axis).}
\label{fig:byteleak}
\end{figure}

\subsection{Determining the Leakage Source}
\label{sec:source-of-leakage}

\parhead{Flushing the MDS Buffers.}
While the \verw instruction is now used by Intel as a defense against MDS attacks, the ability to overwrite the contents of MDS-affected buffers is also helpful in determining the source of the leakage observed by \attackname.
More specifically, our attacker issues the \verw instruction after evicting from the cache, but before executing the leakage primitive.
When the victim and attacker execute sequentially on the same \thread, this completely
removes the signal.
Thus, we conclude that the actual leakage stems from one of the MDS buffers. Next, when we move the \verw instruction before evicting from the cache in our attacker, the attack leaks data from cache lines modified by the victim, but does not leak victim reads.
This supports the hypothesis that the L1-D cache eviction transfers the data into the LFB when it is written back to the L2 cache.

\parhead{Exploiting \verw.}
Ironically, we discovered that issuing the \verw instruction before evicting from the cache significantly improves the signal for victim writes, both in cross-thread and same-thread scenarios.
As the \verw instruction does not require root privileges, we are able to abuse Intel's MDS countermeasures to reduce noise encountered during our attack by having the attacker use \verw to remove unwanted values from the LFB.
We conjecture that this attacker-executed \verw removes all values but the leaked one from the LFB, thereby increasing the probability of the leaked value be successfully recovered by TAA.
To confirm this we run an experiment where we try to leak from writes in the same-thread and cross-thread scenarios, as well as from reads in the cross-thread scenario, both with and without \verw.
Without \verw, we report an actual throughput of 26.57B/s, 2918.33B/s and 343.25B/s for same-thread writes, cross-thread reads and cross-thread writes respectively.
With \verw, we report an actual throughput of 81.45B/s, 1833.93B/s and 2433.97B/s, respectively.

\parhead{Flushing the L1-D Cache.}
To confirm that it is the eviction from the L1-D that causes modified data to transit through the LFB, we try to flushing the L1-D using \ldflush (MSR 0x10b) between the victim access and the cache eviction.
We find that in the same-thread case, this completely removes the signal.
This again supports the hypothesis that evictions of modified data from the L1-D transit through the LFB, where it is leaked by \attackname.

\section{Cross Process Attacks}
\label{sec:cross-process}

To demonstrate the implications of \attackname, we developed multiple proof of concept attacks wherein an unprivileged user process leaks confidential data from another process: recovering AES keys, RSA keys, and the weights of a neural network.
Moreover, in our examples we demonstrate how address selection enables more powerful attacks.
That is, \attackname allows the attacker to select the locations to read in the victim's address space, rather than waiting for data to become available in the LFB.
In particular, unlike ZombieLoad~\cite{zombieload} and RIDL~\cite{ridl}, we can effectively leak random-looking data spanning multiple cache lines. This allows us to lift the known-prefix or known-suffix restriction of~\cite{ridl,zombieload}, which requires prior knowledge of some prefix or suffix of the data to leak.
Indeed, instead of using a known prefix or suffix, we use \attackname to simply read as much data as we can from the L1-D cache.
As we know the location of the data pieces relative to each other, we are able to partially reconstruct a portion of the victim's address space that is located inside the L1-D cache.
Next, we exploit redundancies in the data such as derived AES keys or the relationship between $p, q$ and $n=pq$ for RSA in order to find these inside the reconstructed parts of the victim's memory.

Finally, we also improve on ZombieLoad and RIDL~\cite{zombieload,ridl} by showing attacks with and without \threading.

\parhead{Experimental Setup.}
We run the attacks presented in this section on two machines.
The first is equipped with an Intel Core i7-8665 CPU (Whiskey Lake), running Linux Ubuntu 18.04.3 LTS with a 5.0.0-37 generic kernel.
Our second machine is equipped with an Intel Core i9-9900K (Coffee Lake Refresh, Stepping 13) running Linux Ubuntu 18.04.1 LTS with a 5.3.0-26 generic kernel.
The former machine uses microcode version \texttt{0xca}, while the latter uses \texttt{0xb8}.

\subsection{Recovering AES Keys}

\begin{figure}[b]
\centering
\includegraphics[width=.6\linewidth]{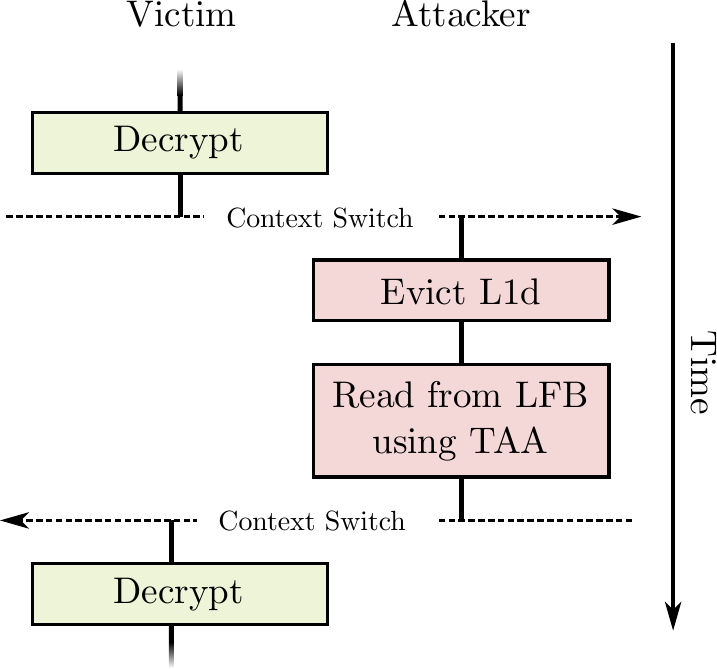}
\caption{%
After decrypting, the victim writes the plaintext, bringing it into the L1-D cache.
The attacker can then evict it from the L1-D cache and use TAA to read it from the LFB.}
\label{fig:sync}
\end{figure}

\parhead{Same-Thread Leakage.}
Our cross-process attack aims to leak plaintext message and key material from an AES decryption operation.
To that aim, we constructed a victim process that repeatedly decrypts an encrypted message, followed by issuing the \yield system call.
The attacking process runs sequentially on the same \thread, and repeatedly calls \yield to allow the victim to run and decrypt the ciphertext.
After the victim finishes running, the attacker evicts the set of interest from the L1-D cache into the LFB.
The attacker then uses TAA to sample the decrypted message from the LFB; see \cref{fig:sync} for an illustration.
Furthermore, we found that if the buffer holding the plaintext messages shares cache line with other data, we can also sample that data.
For instance, we were able to sample the 128-bit AES key from the LFB when the victim writes the plaintext message to the same 64B cache line as the 128-bit AES key.
Finally, even though we artificially instrumented the victim process to yield the CPU to simplify the synchronization problem,~\cite{gullasch2011cache} demonstrate that this is not a fundamental limitation and can be overcome with an attack on the Linux scheduler.

\parhead{Cross-Thread Leakage.}
We also run our experiment with the victim and attacker running on the same physical core, but different threads and without using \yield in either attacker or victim.
Here, we are not only able to see the decrypted plaintext and AES key, but also the expanded round keys used in each of the AES's rounds.
Unlike the same-thread case, we do not require the assumption that the victim performs a write operation into the 64-byte cache line containing the 128-bit AES key.
Finally, since both the initial AES key and the round keys are laid out consecutively in memory, we can use \attackname's address selection capability to recognize the AES keys from within the leakage data.

\parhead{Locating AES Keys.}
To locate the AES keys inside the leakage without knowing the location of the key data structure inside the victim's memory, we follow the technique of~\cite{halderman2009lest} and consider every consecutive chunk of 128 bits, 192 bits or 256 bits of data as a key candidate.
We then expand the candidate into the AES round keys and check if they match the following chunk of data, up to a certain threshold.
If we find such a series of round keys, we conclude that the candidate is the correct key.

\parhead{Experimental Results.} 
Compared to previous techniques~\cite{ridl,zombieload,medusa}, our attack benefits from \attackname's cache line selection capabilities which removes the need to perform online noise reduction techniques (e.g., the ``Domino-bytes'' method of~\cite{zombieload,medusa}).
In addition, unlike previous works, \attackname has the ability to work with and without \threading and to recover the AES round keys.
By targeting specific cache lines from which to leak, our attack classifies plaintext bytes with 96.8\% accuracy and 128-bit AES keys with 90.0\% accuracy, taking 15 seconds on average to recover a single 64 bytes cache line over ten runs for the same-thread setup.

For our cross-thread setup, we sample data from all 64 cache lines in our online phase at 500 iterations of TAA per byte offset.
This part of the attack takes 76.2 seconds on average over ten runs, and we leak data with a raw throughput of 8.90KiB/s and an actual throughput of 63.39B/s.
Furthermore, we observe 98.34\% of the AES key and round keys, where the initial AES key appears at three different locations for 128-bit, and two different locations for 256-bit providing us with additional redundancy.
We then proceed to locate the AES key in our offline phase, which takes 183.29s on average.

\subsection{Recovering RSA Private Keys}
\label{sec:rsa}

\parhead{Leaking from OpenSSL RSA.}
To attack RSA in OpenSSL, we run a victim that repeatedly decrypts a given ciphertext in a loop.
In our setup, both the victim and the attacker run on different threads on the same physical core, where the attacker samples data from the victim using TAA, without the need for \yield in both attacker and victim.
Within the sampled data, we observe 8 byte chunks of $p$ and $q$, though not in any particular order.
Address selection does not help us in this particular scenario, as we observe these chunks of 8 bytes for all possible 8-byte aligned offsets.
Within the raw dump of the sampled data, we are able to observe all of the chunks of $p$ and $q$, but without address selection, we cannot determine whether the chunks are from $p$ or $q$, or where in $p$ and $q$ they appear.

Notably, we did not observe any data from the other components of the private key (i.e., $d$, $d_p$, and $d_q$).
Inspecting OpenSSL's modular exponentiation code, we find that it requires $p$ and $q$ to be repeatedly loaded into the cache, due to OpenSSL's use of the Chinese Reminder Theorem (CRT).
We thus conjecture that the leakage signal observed from the loadings of $p,q$ dominates any other signal.
Our algorithm for reconstructing $p$ and $q$ from the unordered chunks is described in Appendix~\ref{ref:rsa-alg}.

\parhead{Key Extraction Results.}
For this experiment, we generated 512-bit, 1024-bit, 2048-bit and 4096-bit RSA keys.
We then performed the online phase of our attack, sampling sufficient key data from our victim.
We gathered 100\% of the key data in all cases by sampling from a single cache line for 2048-bit keys and smaller, and from four cache lines for 4096-bit keys.
To sample data at each byte offset, we used 3K iterations for 1024-bit and smaller, and 5K iterations for 2048-bit and larger.
Our online phase took 7.4s, 7.4s, 13s, and 51s for the different key sizes, averaged over 5 runs, compared to Medusa's~\cite{medusa} 7min for 1024-bit.

\parhead{Key Reconstruction.}
Next, in the offline phase, we recovered the RSA private keys from the collected data.
We were able to recover 512-bit, 1024-bit, 2048-bit and 4096-bit RSA private keys from the sampled data in 0.3s, 0.3s, 3.5s and 82.8s on average respectively, with the worst-case performance recorded being 186s.
We confirmed correct private key recovery via the corresponding public key. Finally, we note that \attackname's cleaner leakage signal allows us to improve Medusa's~\cite{medusa} 400 CPU hour result for 1024-bit keys to mere seconds, while also attacking larger 2048 and 4096-bit keys.

\subsection{Attacking Neural Networks}
\label{sec:ml}

To further demonstrate the utility of address selection, we also use \attackname to steal the weights from an artificial neural network (ANN).
We note that these weights are valuable IP for companies that invest resources on training networks, which creates economic incentive for stealing such weights~\cite{zhang2018protecting}.
In this section, we consider an attacker that aims to leak the weights from a propriety victim neural network classifier.

\parhead{Recovering Weights from FANN.}
We demonstrate our weight recovery attack against the popular Fast Artificial Neural Network (FANN) Library.
The victim uses the generic FANN model created by \texttt{fann\_create\_standard()} to repeatedly classify a randomly chosen piece of English text as one of three languages.
On a parallel logical thread on the same physical core as the classifier, the attacking process uses 5K iterations to sample data from each byte offset, without the need for \yield in either victim or attacker.
In this manner, we observe 98.4\% of the weights among the extracted data.
However, the vast amount of raw data that \attackname leaks complicates the process of identifying the network's weights, requiring us to use a number of techniques to clean the noise and identify weights' values.

\parhead{Exercising Address Selection.}
The model has 376 weights, with each weight represented with 32 bits, resulting in a 1504B array.
Since the weights are stored sequentially in an array allocated by \texttt{calloc()}, finding the start of the array reveals the page offsets of all of the weights.
After instrumenting the FANN classifier to reveal the address of the weights' array, we found that the array always starts at a fixed page offset.
Thus, the attacker can find this location in an offline phase, thereby enabling her to specifically target the cache lines containing the weights during the online phase. With the naive approach of simply selecting the 8 byte value that has been seen the greatest number of times for each offset containing a weight, we achieve 63.0\% accuracy for determining the value of each weight.
In Appendix \ref{app:weight} we describe how to improve the accuracy to 96.1\% by exploiting both the weights' storage format and the observation that the weights of a neural network tend to be small.
Crucially, we note that without address selection, the attacker would not be able to map the recovered weights to the neural network model.
As the 1504B weight array spans 23 different cache lines, even if the attacker could accurately identify each weight with 100\% accuracy, she would not be able to determine which weight connects which two neurons.

\parhead{Experimental Results.}
We performed an experiment where we try to leak the weights from a trained neural network.
Our attacker took 40s to run on average over ten runs with a raw throughput of 17.08KiB/s and an actual throughput of 662B/s.
We observed 98.4\% of the weights and recover the weights with top-1, top-3 and top-5 accuracies of 95.2\%, 96.6\% and 96.6\% respectively.

\section{Attacks on Linux Kernel}
\label{sec:attack-kernel}

\attackname can also leak sensitive data from the unmodified Linux kernel, even when \threading is disabled.
We demonstrate how by developing attacks for breaking KASLR and recovering secret kernel stack canaries.

\subsection{Derandomizing Kernel ASLR}
\label{sec:derandomizing-kaslr}

\parhead{KASLR Overview.}
Kernel Address Space Layout Randomization (KASLR) is a defense-in-depth countermeasure to binary exploits.
By randomizing offsets of entire code sections, the kernel impedes control flow redirection attacks, which require knowledge of the location of targeted code pieces.

\parhead{Attacking Kernel ASLR.}
We now show how the cache line selection capabilities of \attackname enable an attacker to reliably leak a kernel function pointer and breach KASLR in under a second.
The attacker binds itself to a single core and repeatedly executes a loop composed of just a \yield followed by TAA.
When \yield returns to the attacker from the kernel, we use TAA to leak stale L1-D data leftovers from the kernel during the context switch.
We first used TAA to leak data from all 64 cache lines at all byte offsets.
Upon inspection, we found that a pointer corresponding to the \hrtick kernel symbol could be consistently recovered from the same cache line at the same byte offset.
We then verified that this location remains static across both reboots and different machines running the same kernel version.

\parhead{Attack Evaluation.}
An attacker can exploit this by first conducting offline analysis, running the attack code on a machine running the same kernel version as the victim.
Then, after learning the location, the attacker can conduct the online attack against the victim; the difference is that the attacker needs only leak the single cache line and eight byte offsets that contain the kernel pointer, as opposed to an entire 4KiB of data.
Thus, the cache-line selection capabilities of \attackname result in a running time of 14 seconds for the offline analysis phase, and under a single second for the online attack phase.

\subsection{Defeating Kernel Stack Canaries}
\parhead{Stack Canaries Overview.}
Stack Canaries~\cite{CowanPMHWBBGWZ98} are another widely deployed defense-in-depth countermeasure to binary exploits.More specifically, these aim to protect against stack-based buffer overflows, where an attacker writes beyond the end of a buffer on the stack and overwrites data used for control flow (e.g. function pointers and return addresses).

\parhead{Extracting Kernel Canary Values.}
We used \attackname to leak the Linux kernel's 64-bit stack canary value, which is shared for all kernel functions running on the same core in the context of the same process.
The attacking code is similar to the KASLR break, but instead of repeatedly calling \yield, we execute a loop with a write to \texttt{/dev/null}, followed by performing TAA to leak from the L1-D cache.
We found three different locations (cache line and byte offset) where the kernel's stack canary can be leaked.
On average, the attack succeeds in 23s when evaluated on an i9-9900K stepping 12 CPU with microcode \texttt{0xca} running Ubuntu 16.04.
To our knowledge, \attackname is the first microarchitectural side-channel that manages to recover stack canaries from the kernel.
This is made possible by the address selection capabilities, as a completely random 64-bit value is extremely difficult to detect without targeting a particular cache line.

\section{Breaking Virtualization}
\label{sec:attack-vm}

Infrastructure-as-a-Service (IaaS) cloud-computer services provide their end-users virtualized system resources, where each tenant runs in a separate VM.
Modern processors support virtualization by means of extensions where the hypervisor can create and manage these VMs that each run their own OS and applications in an isolated environment, analogous to how an OS creates and manages processes.
In this section, we demonstrate that \attackname can break VM isolation, showing how to leak both from the hypervisor as well as VMs that are co-resident on the same physical CPU core.

\parhead{Experimental Setup.}
We ran the attacks presented in this section on an Intel Core i7-8665U (Whiskey Lake) running Linux Ubuntu 18.04.3 LTS with a 5.0.0-37 generic kernel and microcode update \texttt{0xca}.
We used QEMU 2.11.1 with KVM enabled and 1GB hugepages set up on two different threads on the same physical core.

\parhead{Evicting L1-D Cache Sets.}
We perform the same experiment as in \cref{sec:l1d-eviction} to determine the number of loads and stores necessary to evict any L1-D cache set across VMs and to attack the hypervisor across CPU threads.
\cref{fig:vm-eviction} shows that we can successfully leak data from the hypervisor as well as across VMs using 8 loads and 3 to 4 stores against the hypervisor and 10 loads and stores across VMs.

\begin{figure}[t]
\centering
\includegraphics[width=\linewidth]{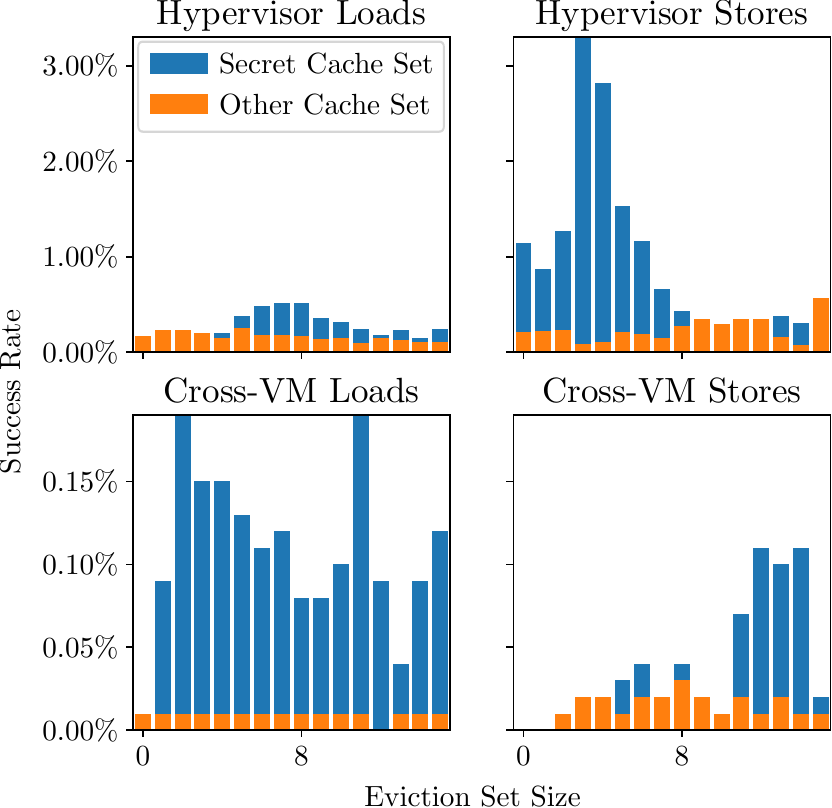}
\caption{%
The number of loads/stores required to evict the L1-D cache sets for loads (left), stores (right), against the hypervisor (top) and across VMs (bottom).}
\label{fig:vm-eviction}
\end{figure}

\parhead{Selecting L1-D Cache Sets.}
After establishing the ideal number of loads and stores required to evict the L1-D cache set, we now proceed with the second experiment as outlined in \cref{sec:l1d-eviction}.
We set up our hypervisor and victim VM to write a secret to every possible cache set, and then try to leak from every possible cache set using our attacker VM.
We present our results in \cref{fig:vmsel}, which clearly shows that we are able to select and evict any L1-D cache set and leak secrets from either the hypervisor or a co-resident victim VM.

\begin{figure}[htb]
\centering
\includegraphics[width=0.99\linewidth]{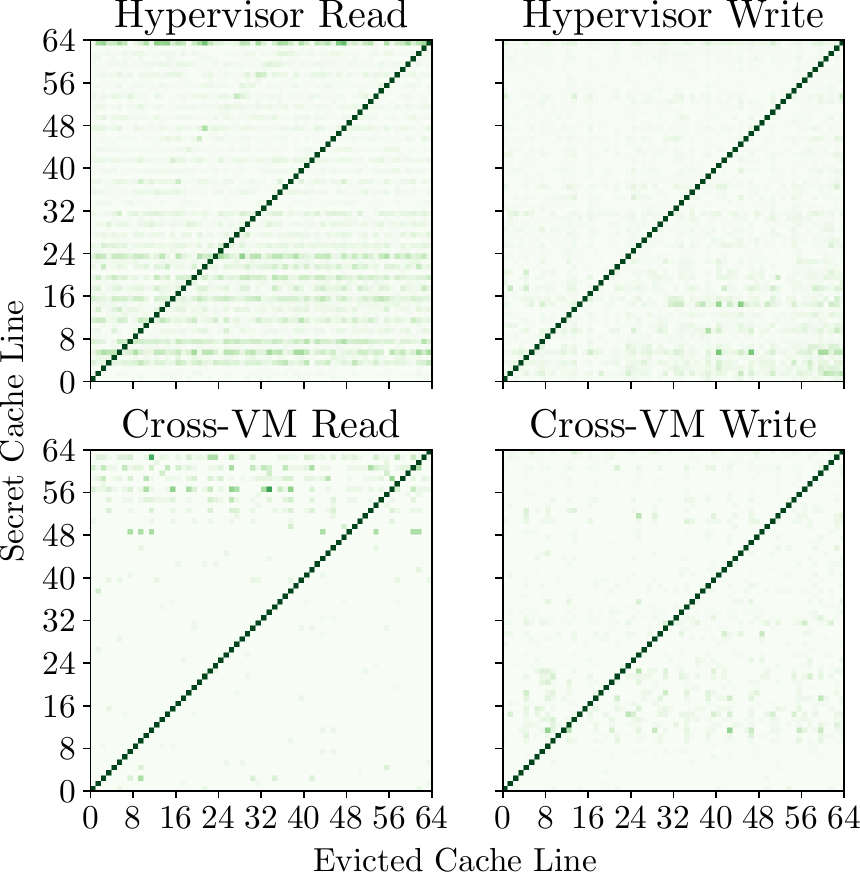}
\caption{%
The hypervisor and victim VM loads/stores a secret from/to every possible cache line (y-axis), while the attacker VM tries to evict every possible cache line (x-axis) to leak it.}
\label{fig:vmsel}
\end{figure}

\subsection{Leaking AES Keys Across VMs}
We run a setup with an attacker and a victim VM across Intel \threads, where the victim is running a program that repeatedly performs AES decryptions using OpenSSL 1.1.1.
The attacker VM evicts the L1-D cache set of interest in an attempt to leak interesting information from the victim process through the line fill buffer.
Once the attacker manages to evict the data from the L1-D cache into the line fill buffer, the attacker uses TAA to sample the AES key.

\parhead{Experimental Results.}
We targeted a specific cache line to leak from, and ran our experiment three times.
For each run, we attempted to leak each key byte 10,000 times.
During all three runs, the bytes corresponding to the victim's AES key were observed during 20 out of the 10,000 attempts.
In order to improve our signal, we run \yield in a loop in an attempt to capture baseline noise that we can later subtract from the AES signal.
Upon subtraction, we were able to recover 75\% of the key bits on average across the three runs.
Finally, as in the case of \cref{sec:cross-process}, it took about 15 seconds on average to leak a single 64-byte cache line.

\subsection{Leaking RSA Keys Across VMs}
We adapted the experiment from \cref{sec:rsa} for stealing RSA private keys across VMs.
We use the same victim as in \cref{sec:rsa}, which runs RSA decryptions in a loop inside a VM.
From within a VM on a parallel \thread running on the same physical core, we use \attackname to sample data from all cache lines.
When repeating the attack from \cref{sec:rsa}, we are able to observe 100\% of the chunks of $p$ and $q$ from the extracted data.
Compared to the cross-process scenario from \cref{sec:rsa}, the VMs introduce a substantial amount of noise, which we overcome by sampling with \attackname over a larger number of iterations.

\parhead{Experimental Results.}
We found that we can extract 100\% of the key data using 5K iterations for 512-bit RSA keys and 10K iterations for 1024-bit RSA keys and larger to sample data per byte offset.
This resulted in an average run time of 11.71s, 23.16s, 23.24s and 24.32 to sample the data for 512-bit, 1024-bit, 2048-bit and 4096-bit RSA keys respectively over ten runs.
In addition, we observed an actual throughput of 1.21KiB/s, 2.15KiB/s, 2.68KiB/s and 5.82KiB/s for the different key sizes.
Our offline phase where we recover the RSA key from our collected data took 2.18s, 2.27s, 33.96s and 95.50s on average for the different key sizes.
Finally, we note that Medusa~\cite{medusa} did not demonstrate any cross-VM data extraction attacks (besides a covert channel), presumably due to noise. 

\subsection{Stealing FANN Weights Across VMs}

We also reproduced the results from \cref{sec:ml} for stealing the weights from FANN.
In our experiment, the same victim from \cref{sec:ml} repeatedly classifies on one VM, while an attacker uses \attackname to sample from an attacking VM running on a parallel Intel \thread on the same physical core.
When using 5,000 iterations of \attackname to leak data from each of the targeted locations, the average run time of the attack is 376.69s with 99.90\% of the weights observed among the recovered data.
We achieve top-1, top-3, and top-5 accuracies of 93.95\%, 96.08\%, and 96.30\% respectively.

\subsection{Breaking Hypervisor ASLR}
Similarly to kernels, hypervisors also deploy ASLR.
To leak any information regarding ASLR from the hypervisor, we first find a controlled way to trap into the hypervisor.
One way of trapping into the hypervisor is by issuing \cpuid from the VM, as the hypervisor hides or represents CPU information in a different way.
We assume an attacker VM with full access over at least a single CPU core with Intel \threading.
On one of the threads, the attacker runs a loop issuing \cpuid, while on the other thread it runs the attacker program.

\parhead{Disambiguating Guest and Host.}
In addition to the hypervisor, our attacker VM is also running its own kernel from which we leak kernel pointers.
In order to disambiguate the kernel pointers we find from actual hypervisor pointers, we simply reboot our VM.
This ensures that the guest kernel has to choose random values again to use for KASLR, while the hypervisor keeps using the same random value.
This allows us to tell apart the pointers we leak from the hypervisor, as the kernel pointers belonging to the attacker's VM are likely to change after a reboot.

\parhead{Hypervisor ASLR Attack Evaluation.}
We first perform an offline phase to determine whether there are static locations from which we can leak hypervisor addresses.
We found that there are indeed various locations that leak a hypervisor pointer to \texttt{x86\_vm\_ops}.
After establishing the fixed locations for a known kernel, we can mount an online attack on the hypervisor.
This reduces the time from roughly 17 minutes in the offline phase to 1.8 seconds.

\section{Breaching SGX Enclaves}
\label{sec:sgx}

Intel's Software Guard Extensions (SGX) is a set of CPU features that offer hardware-backed confidentiality and integrity to user space programs, even in the presence of a root-level adversary.
This enables users to execute a program securely even on a system where the OS and all of the hardware, except for the CPU, are untrusted.
In this section we present attacks for dumping the contents of an SGX enclave, thereby violating SGX's confidentiality guarantees.
Moreover, unlike RIDL~\cite{ridl} and ZombieLoad~\cite{zombieload}, the ability to control which memory address we would like to leak allows us to recover unstructured large secrets, such as images.
Following SGX's threat model, we assume a malicious OS that aims to breach enclave confidentiality.
We also assume that \threading is enabled and that the attacker runs in parallel on the same physical core as the victim enclave.

\parhead{Experimental Setup.}
We ran the attacks in this section on an Intel Core i7-8665 CPU (Whiskey Lake), running Linux Ubuntu 18.04.3 LTS with a 5.0.0-37 generic kernel with microcode version \texttt{0xca}.
This machine is fully mitigated against MDS, meaning that enabling \threading does not violate SGX's security and \threading on Whiskey Lake is considered to be a safe configuration.
Furthermore, the machine has been updated with Intel's latest microcode, which mitigates TAA attacks on SGX by disallowing TSX transactions on logical cores that are co-resident with logical cores running SGX enclaves~\cite{deep-dive:taa}.

\subsection{Reading Enclave Data}

The first building block for attacking SGX with \attackname is to force the victim enclave's data into the L1-D cache.

\parhead{Loading Secret Data into the Cache.}
Even though the malicious kernel cannot directly read the contents of the enclave, the kernel is still responsible for paging the victim enclave's pages using the special SGX instructions \texttt{ewb} and \texttt{eldu}.
Foreshadow~\cite{foreshadow} discovered that by using these instructions, an attacker can load the data into the L1-D cache, even in case the victim enclave is not running at all.
Similarly to Foreshadow~\cite{foreshadow}, we used the \ewb and \eldu instructions to load the victim's decrypted page into the L1-D cache.
See Steps~1 and~2 in \cref{fig:sgx_overview}.

\begin{figure}[b]
\centering
\includegraphics[width=0.99\linewidth]{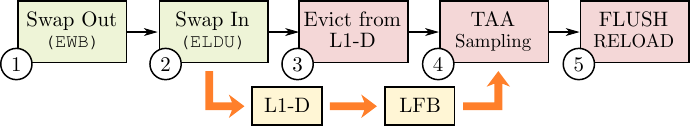}
\vspace{-0.5em}
\caption{%
A schematical overview of how the SGX paging mechanism, in combination with TSX Asynchronous Abort, leaks arbitrary SGX data.
\vspace{-0.7em}}
\label{fig:sgx_overview}
\end{figure}

We improved upon this technique by forcing multiple copies of the plaintext corresponding to the victim's page into the cache.
To achieve this, each time the attacker executes \ewb and \eldu, she allocates a different physical frame for the SGX enclave.
Since writing to different physical addresses puts the data in different cache ways, we were able to fill the entire cache with the victim enclave's secret page, thereby improving the probability of evicting the correct data.
Finally, since the \ewb and \eldu instructions operate at page granularity, an attacker using these instructions can choose which pages to read from.
This gives the attacker more control over the leaked data, compared to the other attacks in this paper, which only have control over the page offset.

\parhead{Reading Secret Enclave Data.}
After loading the secret data into the L1-D (\cref{fig:sgx_overview} Steps~1-2), the attacker can mount a \attackname attack that performs Steps~3-5.
When the attacker evicts the targeted cache line in Step~3, it is evicted into a leaky microarchitectural buffer.
The attacker then leaks the data in the chosen eviction set via TAA (Step~4) and retrieves it using \flushreload in Step~5.
While we chose to demonstrate the attacks in the section against a victim enclave running on a parallel thread, we were also able to observe data leakage in the sequential model.
Even with \threading disabled, we can exercise address selection to read enclave data that remains in the L1-D cache after the enclave exits.

\parhead{Bypassing TAA Countermeasures.}
We are able to utilize TSX for this attack despite Intel's mitigation for preventing TSX and SGX simultaneously running on the same core~\cite{deep-dive:taa}.
This is likely because the L1 is not being flushed after the enclave finishes running.
Afterwards, once TSX is again enabled, the attacker can evict the targeted data from the L1-D into the LFB, and then perform TAA to read the data.

\parhead{SGX Image Extraction.}
In order to quantify our leakage from SGX, we set up an SGX enclave that contains a picture of the Mona Lisa, and use \attackname to leak and reconstruct the picture.
As the image we are trying to extract is 128 by 194 pixels, it spans multiple pages.
Thus, we use the aforementioned \texttt{ewb} and \texttt{eldu} technique on each image page individually, and use address selection in order to leak unstructured pixel data from the entire page.
We sampled the image data from the SGX enclave five times.
For each byte offset, we used 2.5K TAA iterations to sample data, resulting in a run time of 7.75s per cache line and 496s per page on average.
In each such run, we observed 36\% of the image data on average and an actual throughput of 24.54B/s.
We combined the data from our runs observing 71\% of the image.

\parhead{Image Reconstruction.}
We now reconstruct the picture of the Mona Lisa from our collected data.
First, since we have address selection capabilities, we are able to collect all the candidates for every pixel from our sampled data.
Then we calculate a score for each candidate based on the candidates of the neighbouring pixels using a naïve distance function: $(r_1 \cdot r_2)^2 + (g_1 \cdot g_2)^2 + (b_1 \cdot b_2)^2$.
Finally, we sort the candidates based on the smallest score first and select the first candidate as the actual pixel to output in the resulting image.
The offline phase took 8.39s to reconstruct the image, which can be seen on the right in \cref{fig:mona-lisa}.
To reconstruct the the Mona Lisa from our collected data, we first use our address selection capabilities to obtain all the candidates for every pixel from our sampled data.
Then we calculate a score for each candidate based on the candidates for neighboring pixels using a distance function: $(r_1 \cdot r_2)^2 + (g_1 \cdot g_2)^2 + (b_1 \cdot b_2)^2$.
Finally, we sort the candidates select the candidate with the smallest score as the actual pixel value.
The offline phase took 9s to reconstruct the image, which can be seen in \cref{fig:mona-lisa}(right).

\begin{figure}[htb]
\centering
\begin{subfigure}{.5\linewidth}
\centering
\includegraphics[width=.90\linewidth]{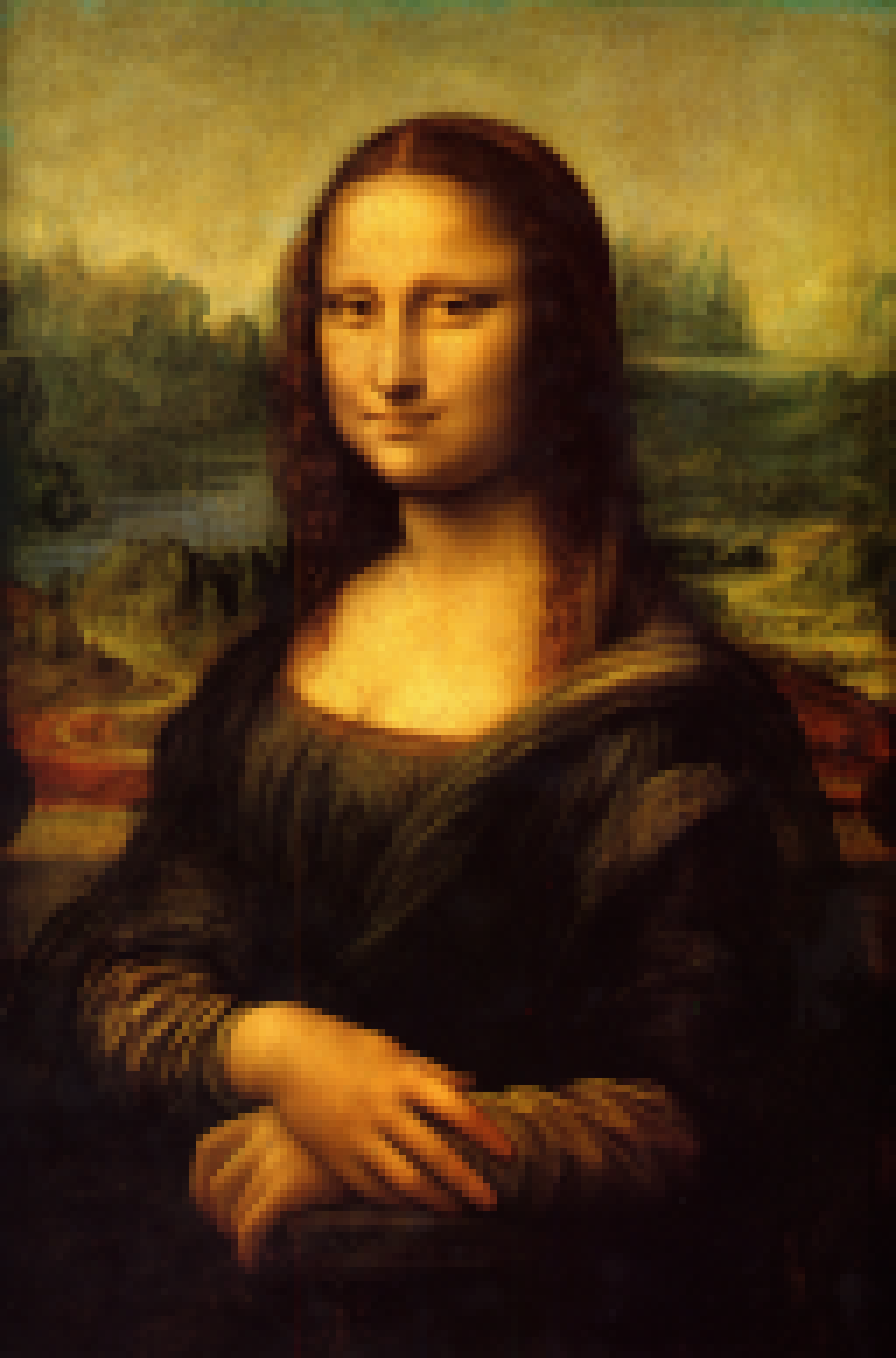}
\end{subfigure}%
\begin{subfigure}{.5\linewidth}
\centering
\includegraphics[width=.90\linewidth]{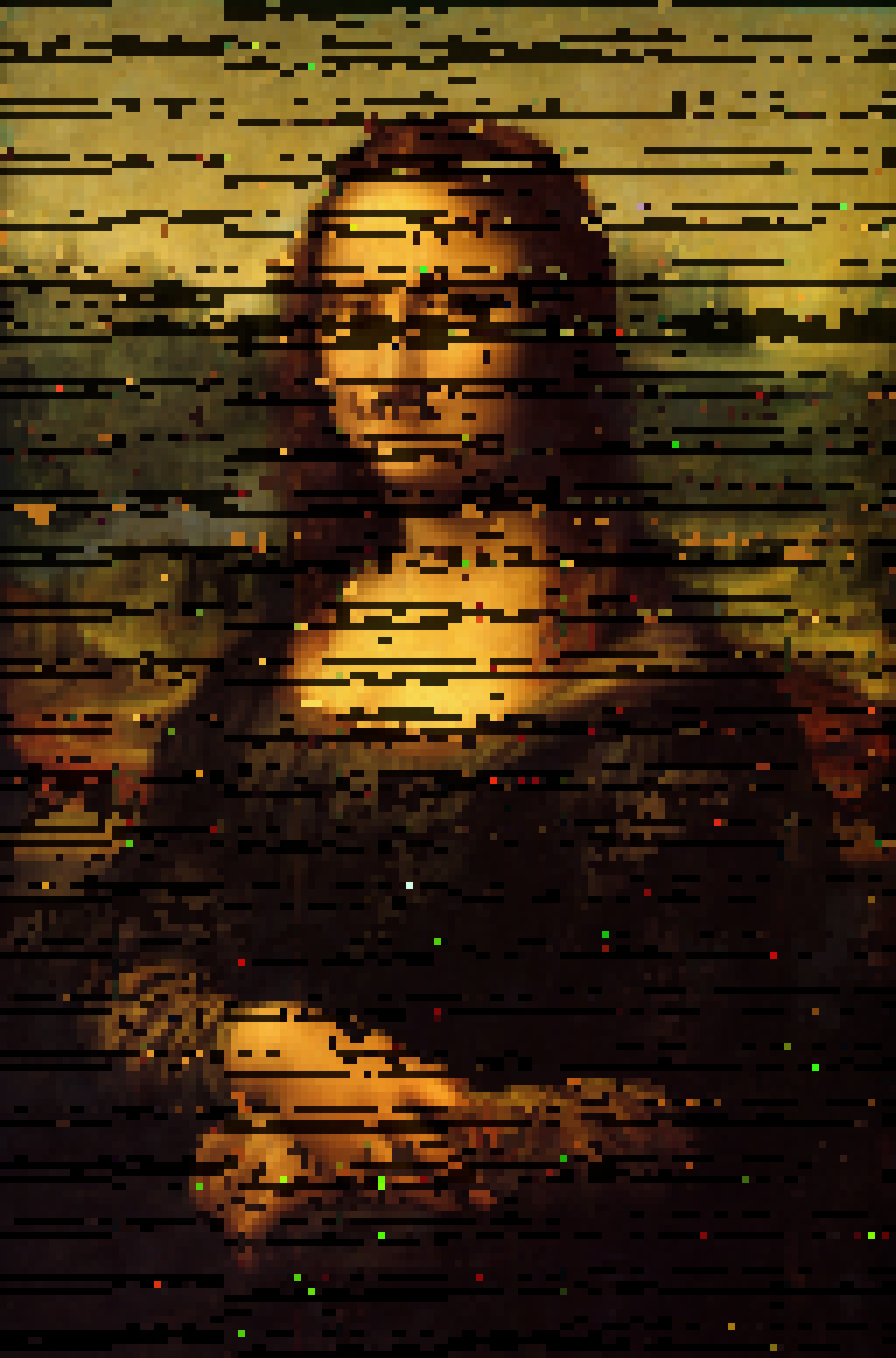}
\end{subfigure}
\vspace{-0.25em}
\caption{%
On the left the original Mona Lisa picture (128x194) and on the right the Mona Lisa picture recovered from an SGX enclave on the Intel Core i7-8665U.}
\label{fig:mona-lisa}
\end{figure}

\subsection{Extracting the SGX EPID Key}
Trust in the SGX ecosystem is rooted in the \emph{Enhanced Privacy ID} (EPID) key, where compromising a single EPID key breaches the entire SGX ecosystem's security.
Thus, this key is available only to enclaves written and signed by Intel.
It is stored as a normal file, but encrypted using \emph{seal keys} that are only available to Intel's quoting and provisional enclaves.

\parhead{EPID Key Extraction in Debug Mode.}
We begin the process of recovering EPID keys by compiling and self-signing Intel's quoting enclave, running it with debuging EPID keys.
We then recovered the sealing key used to seal the file holding the debugging EPID keys and subsequently used it to decrypt the debugging EPID key.
To extract the sealing key, we used a controlled-channel attack~\cite{xu2015controlled}, pausing the quoting enclave after it has loaded the seal key into memory.
After this point, the enclave never resumes execution and is permanently stopped.
After stopping the enclave with the sealing key in memory, we use the technique from \cref{sec:sgx} to repeatedly swap in and out the page containing the sealing key, extracting it using \attackname.
This stage takes about 1.5 minutes.
Due to noise, we see on average 4.5 candidates per key byte, where the key is 16 bytes.
This leaves 747K candidates to brute force.
Because the key is sealed using AES-GCM, we can identify the correct key during the offline brute force phase by comparing against the GCM authentication tags.
We brute-forced the sealing key in 5 seconds, successfully decrypting the file holding the debugging EPID keys.

\parhead{Bypassing Software Defenses.}
We note that the above attack, as well as the attacks in \cref{sec:sgx}, do not require the victim enclave to execute any specific access pattern to the key, or even run at all after loading the key into memory.
Thus, \attackname must be mitigated in hardware, as there is nothing an enclave can do to protect its secrets from being extracted.
In particular, our attack bypasses all existing software mitigations for side channels, such as constant-time coding, detecting page-faults~\cite{shih2017t, chen2018racing, oleksenko2018varys}, and others \cite{fu2017s,sasy2017zerotrace}.

\parhead{Comparison to State-of-the-Art.}
Our breach of an SGX enclave on this particular machine exemplifies how \attackname's advancement over the state of the art in transient-execution attacks enables it to compromise a system that is resistant to previously known attacks.
The i7-8665U (Whiskey Lake) contains in-silicon Foreshadow~\cite{foreshadow} mitigations, which prevent an attacker from directly leaking from the L1-D cache.
Fallout~\cite{fallout} cannot target SGX, as the store-buffer is flushed upon swapping to the enclave's page tables.
Finally, RIDL~\cite{ridl} and ZombieLoad~\cite{zombieload} are mitigated by disallowing TSX and SGX in parallel~\cite{deep-dive:taa}, leaving CacheOut as the only technique for EPID key extraction.

\parhead{Attacking Production Enclaves.}
While the above demonstrates the theoretical feasibility of extracting the CPU's EPID key, we did use a version of the quoting enclave which was self-compiled and self-signed.
As such, this version is unable to access the machine's actual attestation keys, thus preventing us from extracting them.
The reason we made this choice is that at the time of writing, Whiskey Lake machines have an issue with their internal GPU, which allows attackers to leak information from within SGX enclaves~\cite{sa-00219}.
As Whiskey Lake is a laptop architecture, it is impossible to disable the internal GPU, which results in these machines being unable to receive a trusted SGX status and production attestation keys.
Being unable to configure the machine in a state trusted by Intel, we have resorted to extracting the EPID sealing key from the quoting enclave that we compiled and signed ourselves, using the official Intel-provided source code.

\parhead{SGAxe: How SGX Fails in Practice.}
However, in a follow-up work~\cite{sgaxe}, we demonstrate the breach of the SGX ecosystem by extracting production EPID attestation keys from an older Coffee Lake Refresh based desktop, which we were able to configure to a trusted state using an external GPU.
In particular, SGAxe~\cite{sgaxe} demonstrates the extraction of the machine's production attestation keys on a fully updated and trusted machine, defeating recent side-channel countermeasures such as LVI~\cite{lvi} and PlunderVolt~\cite{plundervolt} 

\section{Mitigations}
\label{sec:mitigations}

We now discuss various ways to mitigate \attackname: disabling \threading, flushing the L1-D cache, disabling TSX and microcode updates by Intel.

\parhead{Disabling Hyper-Threading.}
Similar to MDS, \attackname works best when the attacker and victim run in parallel on two threads on the same physical core.
However, as \attackname is also effective in the scenario without \threading where attacker and victim run on the same CPU thread, disabling \threading makes the attack difficult but not impossible (see \cref{sec:cross-process},~\ref{sec:attack-kernel}, and~\ref{sec:attack-vm}).
Finally, as disabling \threading carries a significant performance overhead, we do not recommend this countermeasure for mitigating \attackname.

\parhead{Flushing the L1-D cache.}
As discussed in \cref{sec:source-of-leakage}, \attackname leaks information from the L1-D cache.
Thus, one might attempt to flush the L1-D and LFB on security domain changes, in an attempt to eliminate the source of the signal.
Unfortunately, L1-D cache flushing adds significant overhead and only covers the case without \threading, as leaving \threading enabled means that \attackname will be able to leak data from the L1-D as the victim accesses it.
Thus, given the cost of implementing both of these countermeasures, we do not recommend deploying them for mitigating \attackname.

\parhead{Disabling TSX on New Hardware.}
To address TAA~\cite{deep-dive:taa} on the newest platforms released after Q4 of 2018 (i.e., after  Coffee Lake Refresh), Intel released a series of microcode updates between September and December 2019 that disable transactional memory.
These microcode updates introduce \tsxctrl (MSR 0x122), where the first bit in the MSR disables TSX, and the second bit disables CPUID enumeration for TSX capability.
Concurrent to our work and after our disclosure, OS vendors started disabling TSX by default on all Intel machines released after Q4 of 2018.
We note that, however, this mitigation is only partial as a malicious operating system can always re-enable TSX and use \attackname to leak data from SGX enclaves while bypassing Intel's SGX countermeasures for TAA (as we demonstrated in \cref{sec:sgx}).
Thus, at present SGX remains vulnerable.

\parhead{Disabling TSX on Older Hardware.}
We note however that the vast majority of Intel machines currently deployed were released before Q4 2018. For those machines, Intel started rolling out microcode updates to address CPU errata regarding TSX~\cite{tsx-force-abort}, allowing operating systems to disable TSX by making transactions always abort.
However, at the time of writing this behavior is not enabled by default, leaving the majority of deployed Intel CPUs exposed to \attackname.
Given that TSX is not widely used, we recommend to disable TSX by default on these CPUs as well.
Finally, we note that TSX must be disabled on the microarchitectural level, including during transient and speculative execution, as opposed to aborting the TSX transaction after speculation has occurred.

\parhead{SGX Security.}
As we show in \cref{sec:sgx}, a malicious OS can always re-enable TSX and subsequently use \attackname in order to dump the enclave's contents.
While SGX is insecure at present, we recommend that future microcode updates declare TSX to be unsafe in combination with SGX on current machines, and to flush the L1-D every time TSX is enabled.

\parhead{Microcode Updates.}
Intel's security advisory~\cite{l1des} indicates that \attackname (called L1DES in Intel's terminology) will be mitigated via additional microcode updates.
These are expected to be available on June 9th, 2020, with preview versions supplied by Intel indeed showing a successful mitigation of \attackname.
In private communication, Intel further indicated that mitigating the new data path between L1-D evictions and the LFB discovered by this work is done by adjusting internal CPU timing, preventing the leakage exploited by \attackname.
We recommend that affected users install these, especially on older machines that do not disable TSX by default.

\section{Conclusion}

\noindent In this paper, we investigated Intel's use of buffer overwriting to mitigate MDS attacks, and found that we could force the victim's data to re-enter microarchitectural buffers even after their contents were overwritten during a transition between security domains.
Using this technique we developed \attackname, a new transient-execution attack that is capable of breaching Intel's buffer overwrite countermeasures, while allowing the attacker to surgically choose exactly what data to leak from the CPU's L1-D cache.
We demonstrated the implications of \attackname by developing attacks breaching confidentiality across a number of security domains, spanning user space, kernel space, and hypervisors.
Furthermore, we also demonstrated that SGX is still insecure, despite the deployment of {MDS} countermeasures.
Finally, \attackname is able to leak data on Intel's Whiskey Lake CPUs, which are resilient to prior MDS attacks.

\parhead{Limitations.}
While we clearly demonstrated the feasibility of \attackname using TSX, we were unable to perform \attackname using other transient-execution attack primitives (e.g., mispredicted branches).
While we acknowledge this limitation, we note that TSX is still enabled on all Intel machines released prior to Q4 2018 and can be re-enabled by a malicious OS in the case of SGX.
Next, the signal for the cross-process, VM and kernel variants of \attackname is noisy, requiring multiple attack iterations for data extraction.
Thus, we leave it to future work to demonstrate \attackname-type leakage without TSX, as well as improving that attack's signal-to-noise ratio.
Finally, \attackname is only able to leak data located inside the CPU's L1-D cache, leaving other levels of the memory hierarchy out of reach. As L3 caches are often shared between physical cores, exploring techniques for reading L3-data is an important research problem with many immediate security implications.

\section{Acknowledgments}

This research was supported by the Defense Advanced Research Projects Agency (DARPA) and Air Force Research Laboratory (AFRL) under contracts FA8750-19-C-0531 and HR001 120C0087, by the National Science Foundation under grant CNS-1954712, by an Australian Research Council Discovery Early Career Researcher Award (project number DE200101577), and by generous gifts from Intel and AMD.

\bibliographystyle{IEEEtranSN}
\bibliography{paper}
\FloatBarrier

\begin{appendices}
\section{TSX Asynchronous Abort}
\label{app:taa}

\cref{lst:leak} shows a code example of TAA, where the attacker simply allocates a 4\,KiB page as the leaking source.
She then flushes the cache lines that are about to be used by the TSX transaction, as shown in Lines 5--6.
The transaction then attempts to read from the leak page (Line 10), and then transmits the least significant byte of the value it reads using a \flushreload channel as shown in Lines 11--13.

\begin{listing}[htb]
\small
\begin{minted}{gas}
; %rdi = leak source
; %rsi = FLUSH + RELOAD channel
taa_sample:
	; Cause TSX to abort asynchronously.
	clflush (%rdi)
	clflush (%rsi)

	; Leak a single byte.
	xbegin abort
	movq (%rdi), %rax
	shl $12, %rax
	andq $0xff000, %rax
	movq (%rax, %rsi), %rax
	xend
abort:
	retq

\end{minted}
\vspace{-0.5em}
\caption{the leak primitive using TSX Asynchronous Abort}
\label{lst:leak}
\end{listing}

\section{Recovering p and q} \label{ref:rsa-alg}

With all the chunks making up $p$ and $q$ successfully recovered, the next challenge is to reconstruct $p$ and $q$ such that $N = p \cdot q$.
We assume that the attacker knows the modulus $N$, which is part of the public key.
Then, as observed by \citet{HS09}, $N=p\cdot q$ implies that the low $k$ bits of $N$ are equal to the low $k$ bits of $p \cdot q$.
In order to reconstruct $p$ and $q$, we iteratively recover the the primes 8 bytes at a time as follows, starting from the LSB.

We first iterate over all possible pairs of the 8 byte chunks, and for each pair $(p_0, q_0)$ compute $n_0 \leftarrow p_0 \cdot q_0 $.
If the low 8 bytes of $n_0$ match the least significant 8 byte chunk of $N$, then $p_0$ and $q_0$ are the least significant bytes of $p$ and $q$.
To find the second least significant 8 byte chunks, we again iterate over all pairs and for each pair $(p_1, q_1)$ compute $n_1 \leftarrow (p_1||p_0) \cdot (q_1||q_0)$, where $||$ denotes appending 8 byte chunks.
If the least significant two bytes of $n_1$ are equal to the two low bytes of $N$, then $p_1$ and $q_1$ are the 2nd least significant bytes of $p$ and $q$.
By repeating in this manner for each 8 byte chunk, we can fully recover both $p$ and $q$.

\section{ANN Weight Recovery}
\label{app:weight}

\parhead{Weight Filtering.}
We improve the accuracy of our weight stealing attack by exploiting both the weights' storage format and the observation that the weights of a neural network tend to be small (typically within the range [-1,1]).
The weights are small due to machine learning algorithms using regularization during the training phase, which pushes the weights towards zero in order to prevent both overfitting of the model and the gradient explosion problem, which results in untrainable neural networks.

The weights are stored as 32-bit single-precision floating points, which are specified by the IEEE 754 single-precision floating-point standard to use bit 31 for the sign bit, bits 23-30 for the exponent with a bias of -127, and the remaining 23 bits for the mantissa.
A small value implies that the exponent field will be very near to 127, and despite the 24 bits of precision, this format means that the smallness of the weights result in a very limited set of values for the most significant byte of each weight.
In practice, we find that the MSB does not deviate from 0x40 or 0xc0 by more than 3 for positive and negative weights, respectively.
By rejecting all candidates for weights that do not fit, we improve the accuracy to 93\%.

We further improve the accuracy by observing that the distribution of the frequency of different bytes of noise produced by \attackname is not uniform.
In particular, the values 0x00 and 0xff appear with a far higher frequency than all others.
As such, by penalizing the scores for recovered values that contain 0x00 or 0xff, we improve the accuracy to 96.1\%.

\end{appendices}
\end{document}